# Thermal, Structural, and Optical Analysis of a Balloon-Based Imaging System

**Michael Borden,[a] Derek Lewis,[a] Hared Ochoa,[a] Laura Jones-Wilson,[a] Sara Susca,[a] Michael Porter,[b] Richard Massey,[c] Paul Clark,[c] and Barth Netterfield,[d]**

[a] NASA Jet Propulsion Laboratory, California Institute of Technology, 4800 Oak Grove Drive, Pasadena, CA, USA, 91109

[b] California Institute of Technology, Caltech Optical Observatories, 1200 E California Boulevard, Pasadena, CA, USA, 91101

[c] University of Durham, Centre for Advanced Instrumentation, Durham University, South Road, Durham DH1 3LE, United Kingdom

[d] University of Toronto, Physics department, 60 Saint George Street, Toronto, Ontario M5S1A7

**Abstract**. The Subarcsecond Telescope And BaLloon Experiment, STABLE, is the fine stage of a guidance system for a high-altitude ballooning platform designed to demonstrate subarcsecond pointing stability, over one minute using relatively dim guide stars in the visible spectrum. The STABLE system uses an attitude rate sensor and the motion of the guide star on a detector to control a Fast Steering Mirror in order to stabilize the image. The characteristics of the thermal-optical-mechanical elements in the system directly affect the quality of the point spread function of the guide star on the detector, and so, a series of thermal, structural, and optical models were built to simulate system performance and ultimately inform the final pointing stability predictions. This paper describes the modeling techniques employed in each of these subsystems. The results from those models are discussed in detail, highlighting the development of the worst-case cold and hot cases, the optical metrics generated from the finite element model, and the expected STABLE residual wavefront error and decenter. Finally, the paper concludes with the predicted sensitivities in the STABLE system, which show that thermal deadbanding, structural preloading and self-deflection under different loading conditions, and the speed of individual optical elements were particularly important to the resulting STABLE optical performance.

**Address all correspondence to:** Michael Borden, NASA Jet Propulsion Laboratory, M/S 321-525N. 4800 Oak Grove Dr., Pasadena, CA, USA, 91109; Tel: +1 651-303-2923, Email: mike.b.borden@gmail.com

## 1 Introduction

As astronomers and planetary scientists face shrinking budgets and growing competition for flight opportunities, they are increasingly looking to alternative low-cost platforms that can support science-grade data collection. High-altitude balloons (HABs) are one such platform showing increasing promise for this application. In fact, the planetary science decadal survey for 2013-2022 explicitly called out these platforms for their scientific merit, by suggesting that: "significant planetary work can be done from balloon-based missions flying higher than 45,000 feet… these facilities offer a combination of cost, flexibility, risk tolerance, and support for innovative solutions ideal for the pursuit of certain scientific opportunities."[1] These platforms can reach altitudes that are above much of the earth's atmosphere, offering the large coherence lengths (the propagation distance over which a wave retains its coherence) that provide near-space-like image quality even across spectral bands that are absorbed by the atmosphere and are therefore inaccessible to ground-based observatories. Flights are available at a fraction of the cost of a launch vehicle, and ballooning centers can often recover the payload system so it can be refurbished and reused. These advantages make HABs an attractive solution for certain types of science observations, and have, in turn, fueled the need for the technology to support these observations.

The HAB environment poses several significant technical challenges to data collection that must be addressed before HABs can realize their full potential as science platforms. In particular, thermal and gravity effects combine with the system hardware vibrations to create a complex disturbance environment that makes achieving acceptable optical and pointing performance on the payload challenging. This challenge is even more pronounced for applications with fine pointing needs such as exoplanet observations, galaxy formation studies, and weak lensing/ dark matter and dark energy studies.

The BIT-STABLE (Balloon-borne Imaging Testbed, Subarcsecond Telescope And BaLloon Experiment) project was developed to demonstrate the fine pointing technologies necessary to obtain this kind of science in a HAB environment. This project was developed as a collaboration among the University of Toronto (gondola/coarse stage and ground systems provider), University of Durham, University of Edinburgh (guidance camera provider), and NASA/JPL (fine stage provider). BIT, in the context of the BIT-STABLE project, provides the three-axis attitude control and the coarse stage pointing stability necessary for the fine stage, STABLE, to bring the final system stability to sub-arcsecond levels. BIT performed an experiment in 2015[2] – independent of the BIT-STABLE project – which used a different fine stage and telescope design. Their results suggested that the outer stage was able to stabilize the system to within <0.1° and the inner stage was able to stabilize an image to 0.68" (RMS) over 10-30 minute integrations.[3] Instead, this paper focuses on STABLE, which would use the same BIT outer stage but uses a unique fine stage design.

STABLE addresses two key technology challenges of balloon-borne sub-arcsecond stability platforms: 1) high-bandwidth pointing control loop, and 2) the thermal-structural-optical design that addresses the wide range of expected environmental conditions. This paper specifically describes how STABLE addressed the latter, and presents the thermal-structural-optical modeling, analysis, sensitivity studies, and predicted at-altitude performance of STABLE.

*1.1 State-of-the-Art*

Although a number of missions have aimed to achieve precision pointing on balloons, including SUNRISE[4], Stratoscope II[5], BLAST[6,7], WASP[8], the BIT-STABLE mission has several features that make it a unique solution to the pointing challenges from a HAB. The BIT-STABLE mission is designed to demonstrate 100 milliarcsecond pointing over a 60 second window (1σ) – the level of pointing stability needed to achieve a number of science objectives from a HAB. The guide target for this demonstration is a point source of light (as opposed to an extended source) in the 400-900 nm band with a signal-to-noise (SNR) of 25 (as measured on the STABLE guide detector). These restrictions on the guide target clarify that the mission cannot use the sun for guidance in order to enable night observing, which is how SUNRISE[4] achieved 0.05 arcseconds of pointing stability. Similarly, BIT-STABLE does not rely on infrequent planets or bright stars to enable observing over a wide portion of the sky. The signal-to-noise requirement was determined by scaling from aperture size and the predicted typical guide star brightness values for HAB science missions in development at the time. For the STABLE system, an SNR of 25 is achieved when observing a magnitude 10 star, which can be compared to the magnitude 5-7 stars used for guiding the Stratoscope II mission[5] (which also achieved approximately 0.05 arcseconds of pointing stability). BLAST[6,9], the predecessor for BIT provided coarser pointing than was intended to be achieved by the BIT-STABLE mission, although its disturbance profile was the basis for the development of the BIT-STABLE control loop. The WASP[8] system demonstrated their pointing 5 times on a system with different mass properties. Their latest flight included a payload comparable to STABLE (a 0.5m telescope) and their performance was roughly 4 times worse than STABLE: 0.47 arc sec RMS pitch, 0.39 arc-sec RMS yaw).

*1.2 BIT-STABLE Mission Overview*

Figure 1 shows the main elements of the BIT-STABLE mission architecture. The high-altitude balloon serves as the launch vehicle and the telecommunication relay with the ground, and mechanically connects to the BIT gondola. The BIT gondola, based on a heritage design from the BLAST[6,9] mission, contains all of the batteries, command and telemetry interfacing with the ground, and the coarse pointing stage. This pointing system, consisting of gimbal motors and reaction wheels for actuators, and encoders, star trackers, and gyros for sensors, is designed to lock on to celestial targets and maintain pointing stability of 2 arcseconds (1-σ) over at least two minutes via a series of actuated frames that move in azimuth, roll, and elevation. Connected to the inner frame in the gondola, the STABLE payload consists of the telescope and optical system as well as a power distribution unit, an on-board computer, and a fine pointing system. This stage of the pointing control has a fast steering mirror (a small, piezo-actuated flat optic) as an actuator, and uses a guide camera and an attitude rate sensor for its sensors. As a technology demonstration mission, the BIT-STABLE system does not include a science camera, although plans were made to use a similar system augmented with a science camera for subsequent flights.

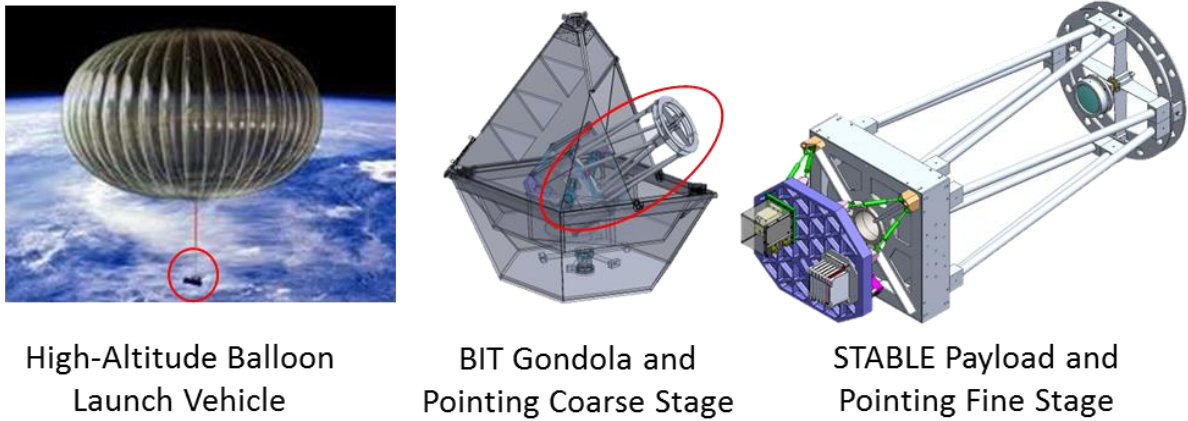

High-Altitude Balloon
Launch Vehicle

BIT Gondola and
Pointing Coarse Stage

STABLE Payload and
Pointing Fine Stage

**Figure 1.** The BIT-STABLE mission architecture.

BIT-STABLE is designed for a single 24-hour flight, with the 8-hour technology demonstration phase of the flight occurring at night to enable a variety of point-source targets at the desired SNR. BIT-STABLE can launch from any one of three launch sites: Kiruna, Sweden; Timmins, Ontario, Canada; and Fort Sumner, New Mexico, United States. Most launch opportunities at these sites are in the spring and fall while limited opportunities exist in winter. The target altitude for the mission is 35 km, although higher altitudes are better for astronomical observing and are less stressing thermally. For the STABLE project, altitudes between 30 km and 40 km are considered in assessing observational thermal performance.

Operationally, the BIT-STABLE mission would observe a star of the target magnitude for approximately 10 minutes. The BIT gondola identifies the appropriate part of the sky to observe and provides the coarse pointing to the target, and then commands the STABLE payload to engage the fine loop. When the observation is complete, the BIT gondola then points to the next target and repeats the sequence. This observing time is planned to last a minimum of eight hours, after which the BIT-STABLE hardware would be released from the balloon and recovered by the launch providers. Figure 2 shows the planned concept of operations for the BIT-STABLE mission.

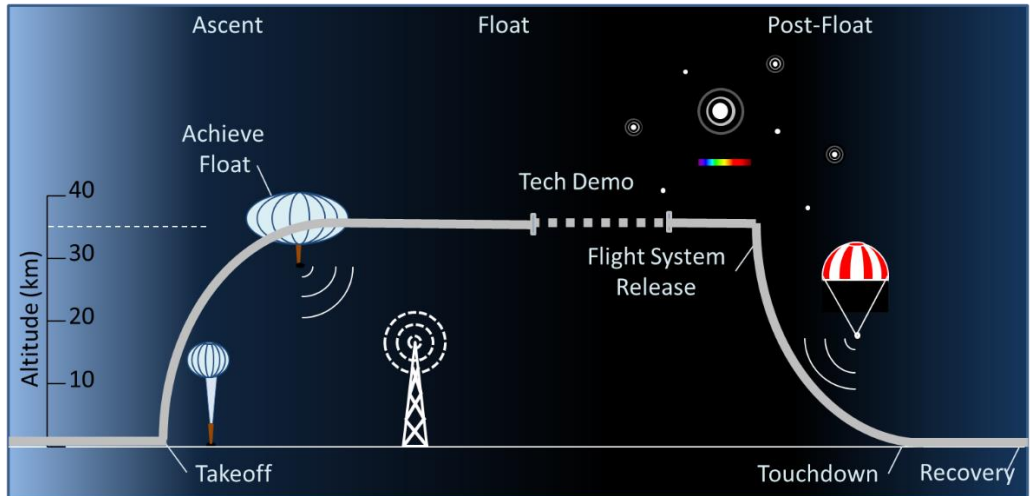

**Figure 2.** The BIT-STABLE mission concept of operations. Note that the time of day of the takeoff and touchdown varies by launch facility – a dusk launch as shown is typically for a Timmons flight, but a dawn launch is associated with a Ft. Sumner flight. This figure is not intended to suggest that a Ft. Sumner flight is precluded from the BIT-STABLE CONOPS.

### 1.3 STABLE System Overview

*System Resources*

In the as-built system, shown in Figure 3, the total mass of the STABLE payload (up to, but not including, the BIT gondola's inner frame) is 155.35 kg, and the total predicted power consumption is 152 W (average), up to approximately 700 W (peak). Over the 24 hour notional mission concept of operations, the predicted energy consumption of the STABLE system is 2747 W-hr. (Note that different launch sites have different total flight durations, but 24 hours represents the maximum total duration expected of the potential BIT-STABLE launch sites.)

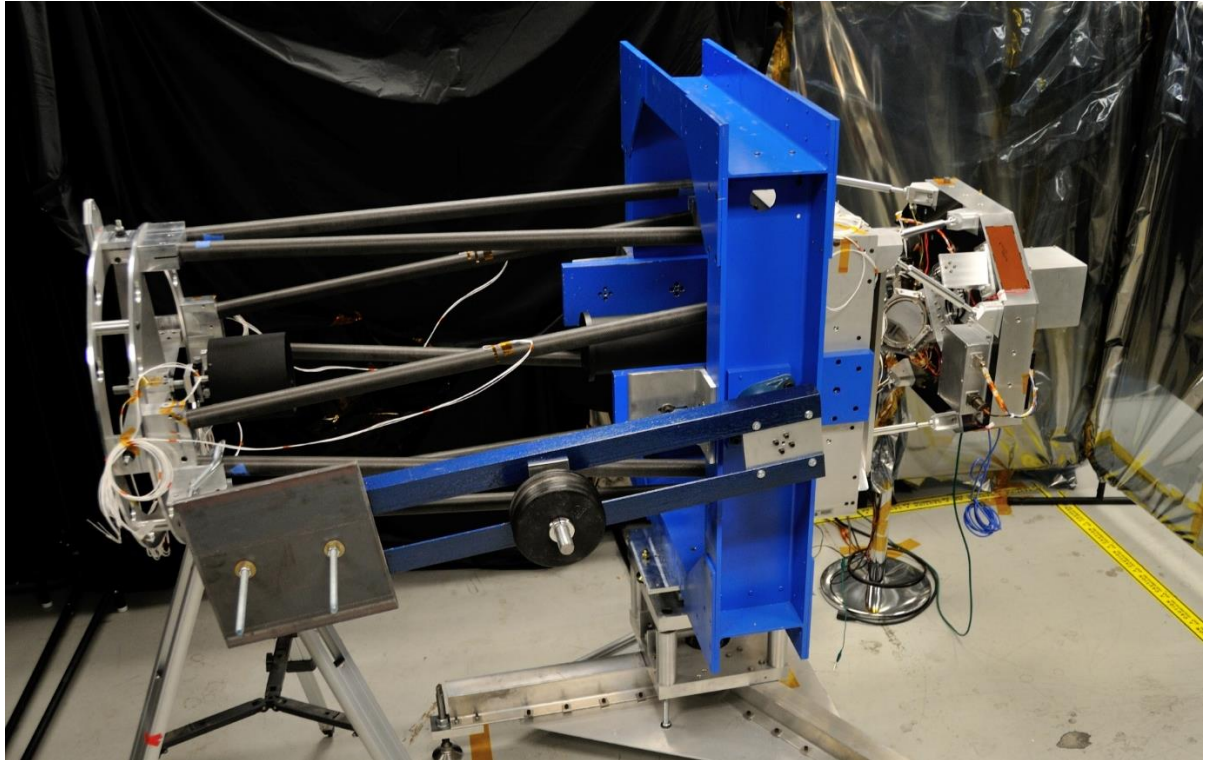

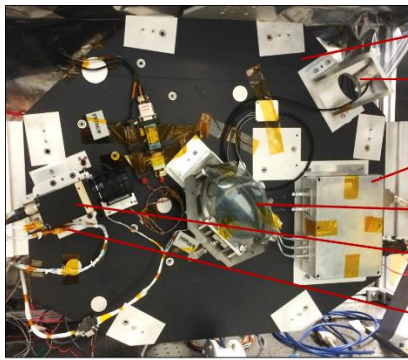

**Black Kapton Tape (THR)**
JPL In-House

**Fast Steering Mirror (FSM)**
Physik Instrumente Actuator S-330
Edmund Optics Mirror 64-019 Custom

**FSM Electronics (FSM ELEC)**
JPL In-House

**Fold Mirror (FDM)**
JPL In-House

**Fine Guidance Camera (CAM)**
Basler A2320

**Refocusing Stage (RFS)**
Zaber Technologies Inc.
T-LSM050A-SV1

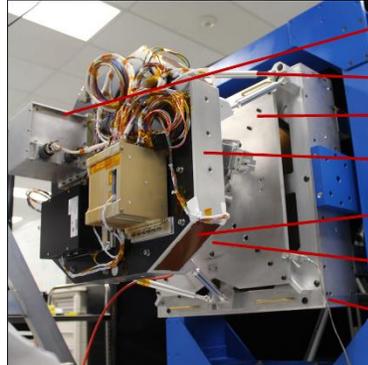

**Attitude Rate Sensor (ARS)**
Applied Technology Associates
Multi-Axis ARS Dynapak

**Bipod Mount to Telescope**
JPL In-House

**Telescope Stiffener Plate**
JPL In-House

**Optical Bench Assembly (OBA)**
JPL In-House

**Heater Assembly 1&2**
MINCO Polyimide Thermofoil
Honeywell 3200 Series Thermostats

**Temperature Sensors (PRTs)**
Honeywell HRTS PRTs
Ohmite 43F7K5E 7.5 kilo-ohm Resistors

**Telescope**
Equinox Interscience
0.5m Ritchey-Chretien

**Figure 3**. STABLE hardware components, including both sides of the Integrated Optical Bench Assembly

*System Hardware*

STABLE is composed of two main elements: the Integrated Optical Bench Assembly (IOBA), which is the optical bench in the rear of the telescope, and the telescope consisting of a primary and secondary mirror pair as shown in Figure 4. The eight-sided IOBA is a custom in-house JPL design that serves as the main mechanical interface and precision metering structure for all of STABLE's electronics, sensors, actuators, and back end optical train.

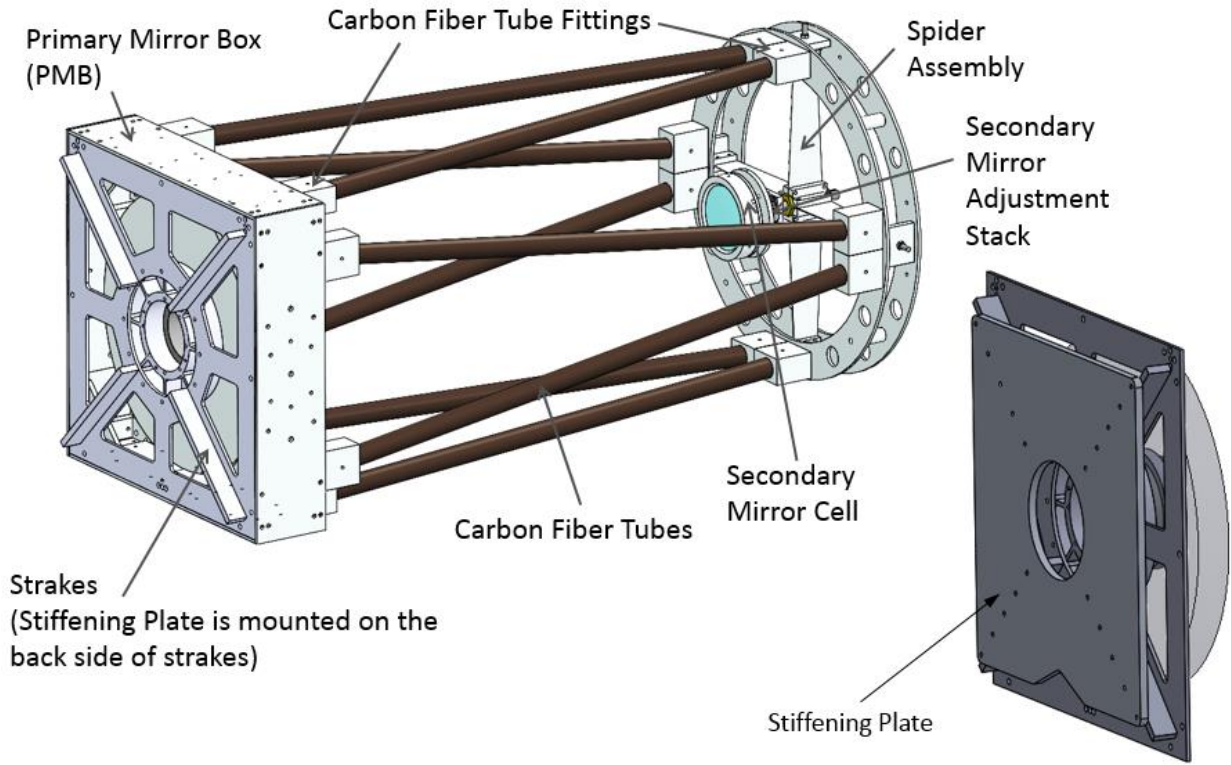

**Figure 4.** STABLE telescope components and modifications

The telescope, built by Equinox Interscience, is based on the design of a ground-based telescope built by the same vendor. The structural components of the telescope are shown in Figure 4. This includes the primary mirror box (PMB), which houses the primary mirror and its mount. It also serves as the interface for the secondary mirror assembly, which includes carbon fiber tubes, a spider assembly, and the secondary mirror mount.

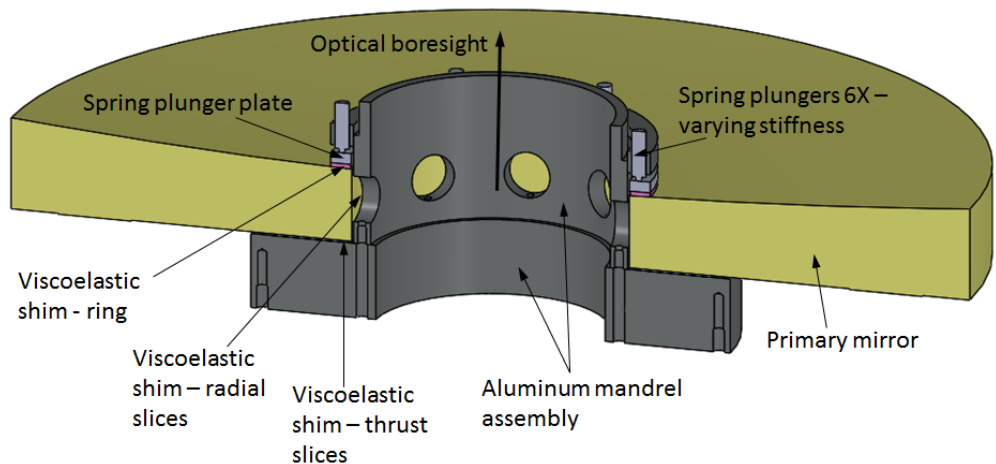

**Figure 5**. STABLE primary mirror mount design

As shown in Figure 5, the telescope's primary mirror mount relies on a clamping force to restrain the mirror axially around its center bore. The primary mirror mounting solution underwent a number of design iterations, using the results of analysis to guide the design decision-making process, but the final design incorporates six spring plungers into the mandrel assembly, along with viscoelastic shims. This design ensures that as the mirror and mount cool, the spring plungers compress with only a small increase in axial preload. To support the primary mirror radially, an additional set viscoelastic shims are installed between the outer diameter of the mandrel and inner diameter of the mirror center bore, as illustrated in Figure 5.

For the secondary mirror mount, a stainless steel mirror cell is used which supports the mirror using nine radial RTV bond pads. This design is shown in Figure 6.

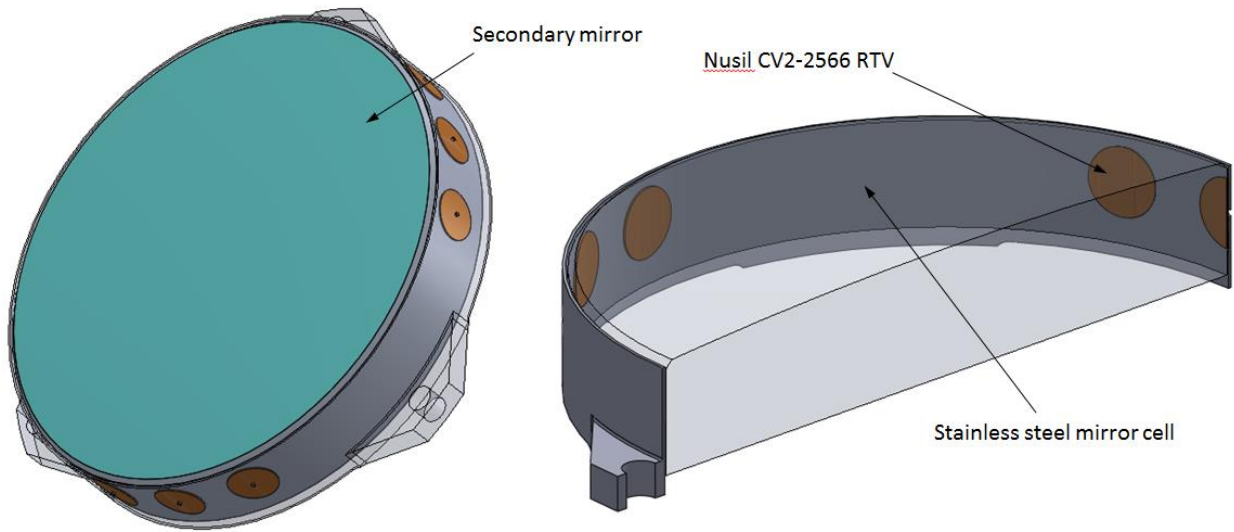

**Figure 6**. STABLE secondary mirror mount design

The STABLE fine guidance camera is a Basler A2320 off-the-shelf CCD with mirolenses and 5.5 µm pixels (0.13 arcseconds on the sky).[10] The unit has a detector of 2336 x 1752 pixels, which corresponds to 4.91 arcminutes x 3.68 arcminutes on the sky and 12.85 x 9.64 mm in physical extent. The STABLE pointing system windows the detector to 100 x 100 pixels (0.21 arcminutes on the sky); of which the gondola 3σ stability predictions would generate motion within a 47.5 pixel radius (6 arcseconds on the sky) over two minutes. The STABLE pointing system then controls the spot stability to within a 3σ motion of 2.4 pixels radius (0.3 arcseconds on the sky).

*Thermal, Structural, and Optical Design*

Thermal systems on balloon missions are often primarily in place to maintain the operating temperatures for the system components. Although STABLE's thermal system does perform this function, its more complex thermal requirements come from the need to maintain the system's mechanical and optical performance. The STABLE thermal design accomplishes both objectives by relying on surface finishes, heaters, and temperature sensors to maintain the system performance in a variety of environmental conditions. The telescope is not actively heated, relying instead on an athermal mechanical design to maintain optical performance over the temperature ranges expected over the mission. The IOBA, on the other hand, has four heaters controlled by two thermostats that maintain the bench temperature between 2 and 8 degrees

Celsius. Ten PRT temperature sensors are located across the system and their values are transmitted in downlinked telemetry, which enables ground operators to reconstruct the as-flown effective focal distance to evaluate system performance.

The STABLE mechanical system is designed to both withstand the load environment of the balloon launch and maintain the relative positioning of the optics and sensors/actuators across different thermal and loading scenarios. STABLE's mechanical design routes the primary load path through the telescope's PMB, which connects to the gondola inner frame by way of a set of bipods. The IOBA is mounted to the telescope PMB by way of kinematic bipods, shown in Figure 3. A number of mechanical features on the telescope also act to maintain the system's optical performance, including a stiffening plate in the rear of the telescope to limit the PMB flexing modes and low coefficient of thermal expansion carbon fiber tubes to limit thermally induced motion between the primary and secondary mirrors.

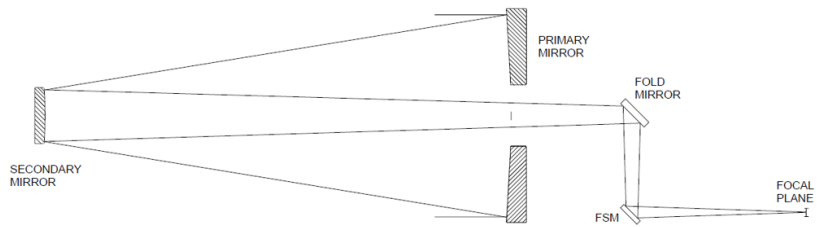

| Telescope Optical Prescription | |
| --- | --- |
| Parameter | Value |
| Primary Mirror Diameter [mm] | 508 |
| Primary Mirror Radius of Curv [mm] | 3048 |
| Primary Mirror Conic Const | -1.01796 |
| Secondary Mirror Diameter [mm] | 123.292 |
| Secondary Mirror Radius of Curv [mm] | -875.856 |
| Secondary Mirror Conic Const | -2.11179 |
| Mirror Separation Distance [mm] | 1160.17 |
| System Focal Length [mm] | 9006.66 |
| System F/# (500mm Aperture) | 18.01 |

**Figure 7.** STABLE optical design and prescription

The STABLE optical design is responsible for projecting the target star onto the STABLE camera detector and limiting the errors in the point spread function (PSF). The F/18, Ritchey-Cretien telescope has the only powered optics in the system: the fold and steering mirrors are both flat, as shown in Figure 7. The F/3 primary mirror is 0.5 m in diameter and made of Zerodur Class 0 with an aluminum coating. This fast mirror makes the spacing between the primary and secondary mirrors highly sensitive: 1 um of spacing change generates 37 um of system focus shift. The secondary is a 12cm mirror made of Zerodur Class 0 with an aluminum coating. The secondary cannot be actuated during flight, although it can be adjusted during ground alignment in tip and tilt, piston, and translation in X and Y. Instead, the STABLE camera is attached to a single-axis translation stage, shown in Figure 8, which moves the detector along the optical axis to a system focus during the flight. The system fold mirror can also be adjusted on the ground in tip and tilt to facilitate alignment. Although the fast steering mirror could be actuated to remove decenter, both in ground alignment and during the flight, the STABLE alignment process was specifically developed to avoid using stroke to solve decenter issues in order to preserve the stroke available to the control system. STABLE's mission does not necessitate tight requirements on pointing accuracy, and so the system is centroided on whichever detector location on which the star is first acquired. STABLE's mission also does not require diffraction-limited observing; rather, the pointing system simply requires a Strehl ratio of greater than 0.6 across the nominal mission scenarios.

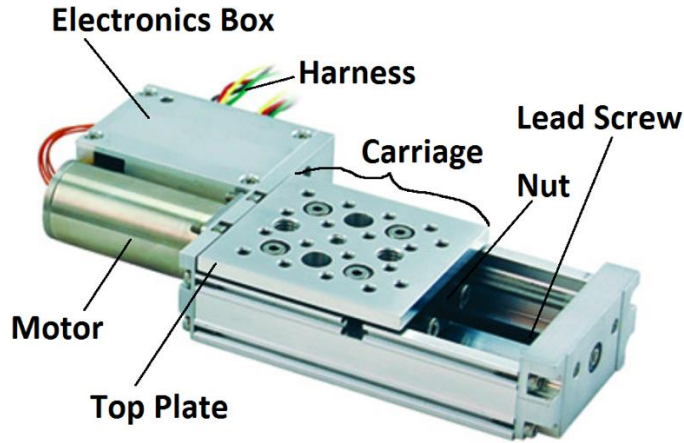

**Figure 8.** Zaber T-LSM050A-SV2 vacuum compatible translation stage

*1.4 Thermal-Optical-Mechanical Analysis Overview*

In order to evaluate the expected system performance over the wide variety of possible environmental conditions, it is critical to understand the system's thermal-mechanical-optical interdependencies. As such, the STABLE team performed an extensive Structural, Thermal, Optical Performance (STOP) modeling effort that informed the design of these critical subsystems and generated the end-to-end performance estimates for the as-built STABLE system. This type of comprehensive analysis is common for space-based systems with a much higher budget. Figure 9 shows the interface products and models associated with the STABLE STOP analysis. This paper describes the thermal, structural, and optical models used in the analysis, explains the results of each of these analyses, and details the sensitivities of the STABLE system that drove the ultimate design and performance in each of these subsystems. The modeling methodology used for the thermal-optical portion of this mission is consistent with the approach used on the CIDRE instrument.[11]

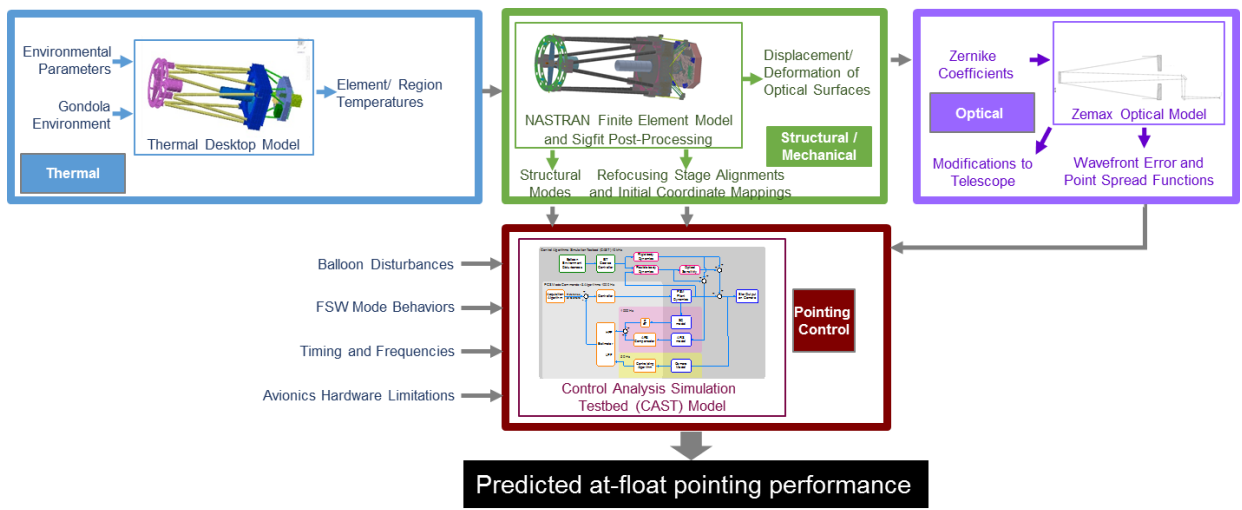

**Figure 9.** STABLE STOP analysis workflow

## 2    Thermal Analysis

### 2.1 Thermal Modeling Approach and Case DefinitionsOverview

Thermal modeling is the first step in the STOP analysis because the bounding thermal cases this analysis produces can profoundly influence the displacement and deformation of the system structure and optical elements.  Multiple inputs are required for the thermal analysis: mechanical design, thermal design, instrument power modes, mass characteristics, and environmental parameters to name a few. Ultimately, the iterative process of design and analysis led to the STABLE telescope configuration used in the final STOP analysis. A Thermal Desktop model captured this configuration along with the additional inputs required for analysis. Then, three fully transient cases (from launch and ascent, to end of flight), were analyzed: two for bounding stacked "worst-case" assumptions, and one with nominal assumptions. Then, discrete points in the worst-case-hot, worst-case-cold, and nominal transient thermal results were selected to span the range of possible telescope performance. Note the telescope was not in thermal equilibrium in any of these instances in time. Additionally, both models were discretized into thermal zones (Figure 10) to reduce the time required to map the thermal and structural models. This reduced the level of effort typically required for mapping the thermal model with the structural model.

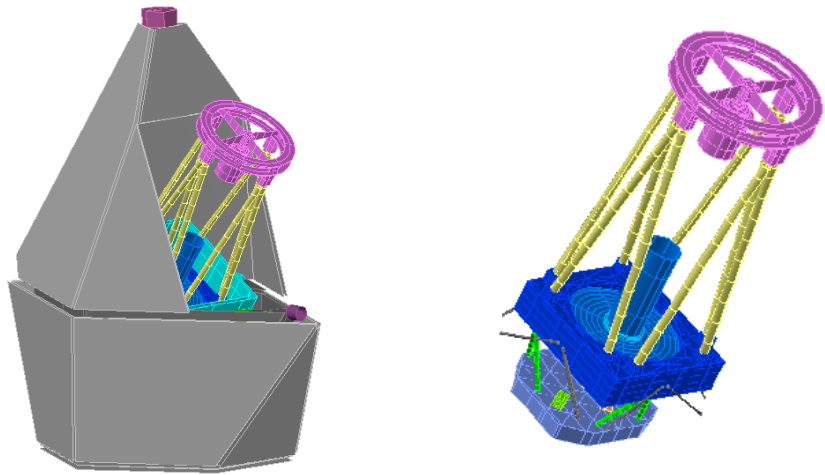

**Figure 10:** Left: BIT-STABLE Thermal Desktop model. Right: Model with the Gondola hidden.

### Case Definition

Given the variety of potential launch sites and the many factors that affect the thermal loads during a mission, there are a myriad of potential thermal cases to consider. STABLE chose to limit the analysis to three thermal transient cases with the key difference between the three cases being the assumed environmental parameters: a stacked worst-case hot assumptions case, stacked worst-case cold assumptions case, and nominal case using the average environmental parameter values of the two worst case. Table 1 provides an overview of the three transient thermal cases.

**Table 1:** Physical conditions describing the hot, cold, and nominal thermal cases

| Parameter | Hot Case | Cold Case | Nominal Case |
|---|---|---|---|
| Launch Time | Dawn | Dusk | Morning |
| Flight Duration | 24 hours | 12 hours | 24 hours |
| Day light | 12 hours | 0 hours | 10 hours |
| Float Altitude | 40 km | 30km | 35km |
| Launch Site | Fort Sumner | Kiruna, SWE | Fort Sumner |

Although these three thermal cases were analyzed as transients with the entire flight duration being modeled, (launch, ascent, observation, and end of flight) only two points in time were used to bound the full STOP analysis: the coldest predicted temperature of the telescope assembly, (occurring at end of the observation phase in the worst-case cold transient case), and the warmest predicted temperatures (occurring at the beginning of the observation phase in the worst case hot transient case. . Hence, these two sets of results from the full transient thermal analyses are mapped to the FEM model to assess the worst-case cold (WCC) and the worst-case hot (WCH) temperature impacts on the mechanical and optical systems. Similarly, in order to evaluate the performance changes over an average flight, two points in the nominal transient results were prepared for the FEM analysis: beginning of night (immediately prior to the start of the technology demonstration phase of the mission) and the end of night (after 8 hours of the technology demonstration phase). These four cases then define the thermal conditions for the entire structural, thermal, optical system performance analysis.

## 2.2 Environmental Parameter Modeling

The STABLE approach to developing the worst-case and nominal thermal conditions involved developing a bounding hot and cold case for each environmental parameter using methods from previous balloon flight projects and publicly available historical data. For simplicity, all potential launch sites and dates were considered in the same pool of data and the thermal cases were developed from the average and bounding cases from that entire pool.

### Air Temperature

The air temperature to be used during the ascend and float portion of the thermal analysis is estimated using monthly radiosonde data made available by the University Of Wyoming Department Of Atmospheric Science.[12] Representative locations close to the potential launch sites were used: Maniwaki, Canada for Timmins; Oland, Sweden for Kiruna; and Albuquerque, NM, US for Fort Sumner. Figure 11 plots all temperature data from the three sites for a six year period – excluding December, January, and February because of the low likelihood of a launch during these months. One important trend to note is the fact that between approximately 25 km and 40 km, the temperature actually increases. Thus, if the balloon achieves a lower altitude than expected, the temperatures are likely to be much colder. The temperature data is discretized into altitude windows corresponding to the different phases during flight: 25km-30km, 30km-35km, and 35km-40km. At each of these windows, the distribution of temperature, shown as histograms in Figure 12 is used to develop an appropriate set of bounding cases.

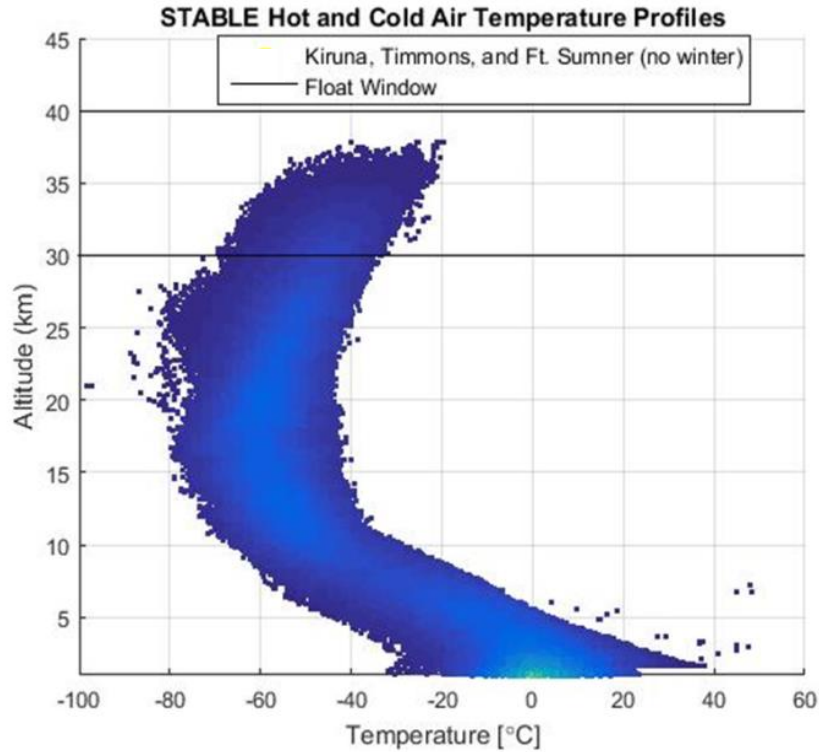

**Figure 11.** Altitude vs. temperature for three launch locations and original air temperature profiles used in the thermal analysis. Darker shade is less frequent data and lighter shade is more frequent data.

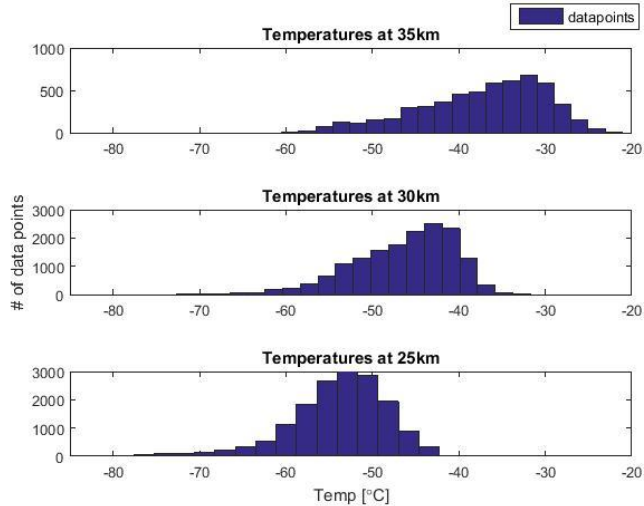

**Figure 12.** Histogram of radiosonde data for 35 km (top), 30 km (middle), and 25 km (bottom)

Note that the number of radiosonde data points for altitudes above 35 km is scarce and it is difficult to characterize the air temperature distribution up to the highest expected floating altitude of 40 km. Also, it is interesting to note that for all three altitude windows the data is left-skewed: extremely cold temperatures, although occurring, are not frequent. The skewed temperature distribution led the project to adopt the air temperature profiles shown in Figure 13. Data within two standard deviations from the mean is used to assess bounding conditions for observation altitudes (30km – 40 km) and only one standard deviation of the data is used for other altitudes (where STABLE is expected to remain for relatively short periods during ascent and descent).

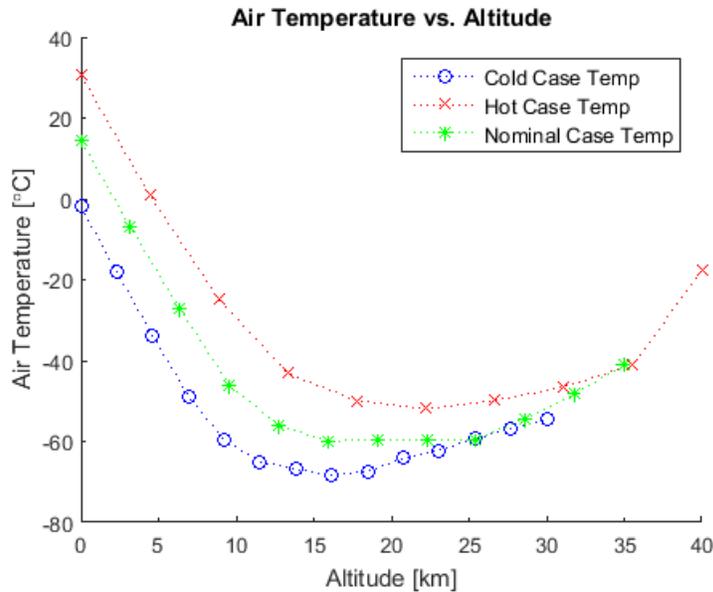

**Figure 13.** STABLE Air Temperature vs. Altitude. These three profiles are used for the three transient thermal cases.

*Forced Convective Heat Transfer Coefficient*

The forced convective heat transfer coefficient, $h_{forced}$, depends on the air velocity, air temperature, surface temperature, and system geometry; hence, it can be difficult to predict. Two NASA balloon flight projects: the Low Density Supersonic Decelerator (LDSD) and the Viking Balloon-Launched Decelerator Test (VLDT) were used to generate the STABLE bounding convection coefficient assumptions. LDSD used values in the range 15 W/(m²-K) to 0.05 W/(m²-K), and VLDT used values in the range 4 W/(m²-K) to 0.6 W/(m²-K).[13,14] These coefficients depend on the air velocity, air temperature, surface temperature, and system geometry. As such, a first-order non-iterative analysis is used to estimate the heat transfer coefficients specifically for the BIT-STABLE system assuming a sphere geometry of the same surface area as the flight hardware. The results fell within the range used of LDSD. Figure 14 shows the bounding coefficients used for the BIT-STABLE mission.

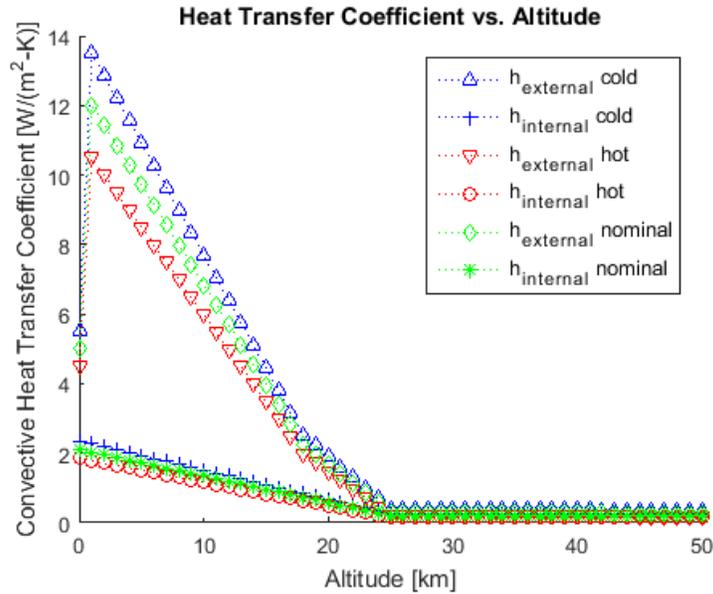

**Figure 14.** STABLE estimated heat transfer coefficients over time

*Infrared Thermal Radiation*

The infrared thermal radiation from the Earth's surface (upward IR) and the surrounding atmosphere (downward sky IR) is modeled following the model proposed by the Scientific Ballooning Handbook.[15] As shown in Figure 15a, STABLE modeled the upward IR temperature as varying linearly with altitude, starting at ground temperature until 12 km, and stabilizing to 20°C below local air temperature. Similarly, as shown in Figure 15b, the sky IR temperatures start from 18°C below ground temperature at zero elevation and linearly decrease until 30 km to the near-space temperature of -245°C.

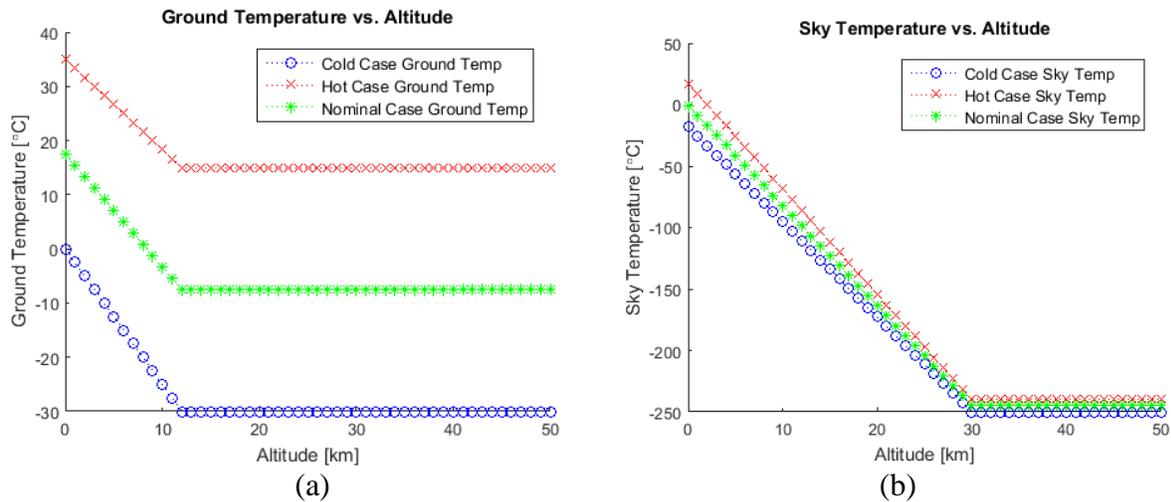

(a)                                                                 (b)

**Figure 15.** STABLE estimated (a) ground temperature and (b) sky temperature over altitude

*Solar Radiation and Albedo*

Solar loads during the day are due to direct solar flux as well as reflected radiation due to the earth's albedo. Direct solar radiation is estimated using an approach similar to the VLDT program to take into account atmosphere attenuation, where the solar flux is a function of altitude, $z$, and the zenith angle, (in itself a function of latitude, longitude, and time of year). Buna and Battley describes in detail the calculation of this atmosphere attenuated solar flux.[13]

The zenith angle is also used to determine the evening end time and morning onset time of astronomical twilight during the possible dates of the flight. Knowing these estimated times helped in filtering out launch dates that would not meet the mission's minimum observational time requirements of eight hours.

The surface albedo is assumed to vary linearly with altitude starting at surfaces of newly paved asphalt (cold case) and new concrete (hot case). The surface albedo at launch altitudes is estimated from data provided by NASA's Clouds and Earth's Radiant Energy System (CERES).[16] The lowest and highest albedos across all three launch sites that were used to determine the appropriate cases for the thermal analysis are shown in Figure 16.

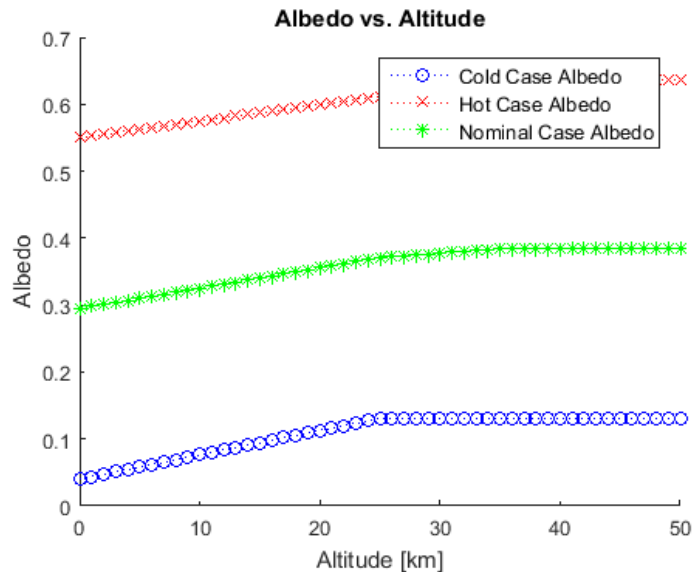

**Figure 16.** STABLE estimated albedo over altitude

## 2.3 Thermal Analysis and Results

These environmental parameters and assumptions, combined with the system design, were used to develop a thermal model of BIT-STABLE in Thermal Desktop. Use of this tool is in alignment with common practice of other balloon flight projects[17,18]. The modeling methodology adapted by STABLE is in family with that of the LDSD thermal analysis and more details can be found in Mastropietro, 2013[14]. The desire to have environmental parameters vary as a function of launch site and altitude precluded the use of the default constants used in Thermal Desktop's Orbital Manager. Instead, for an assumed launch, ascent, float, and landing altitude profile all environmental parameters were converted to time varying inputs for the analysis with the initial environmental conditions defined as ground level conditions and initial telescope temperatures assumed to be room temperature. For example, a time varying boundary node represents the ambient air temperature values in Figure 13 and surfaces fully exposed to the external environment were thermally coupled to this boundary node with a time varying conductor using the external heat transfer coefficient values in Figure 14. Similarly, an arithmetic node was used to represent the air temperature inside the Gondola and all internal surfaces were coupled to this node using the internal heat transfer coefficients shown in Figure 14. Although the thermal model did include a representation of the BIT Gondola, the temperature results of the Gondola were not used in the STOP analysis. The Gondola-to-STABLE thermal interface is isolated with the use of low conductance mounting fixtures and the use of low emissivity finish for the inner surface of the Gondola. A transient thermal analysis was performed across the worst-case hot, worst-case cold, and nominal thermal cases. The thermal model has fewer nodes than the structural model in order to manage the scope of the analysis, so the telescope is divided into simplified isothermal structural groups to enable the thermal nodes to map to an appropriate group of finite element model (FEM) nodes (Figure 17). The primary mirror, being an important element in the optical performance, had more thermal nodes and mapped more closely to the structural nodes because of its complexity and criticality to the optical performance of the STABLE system.

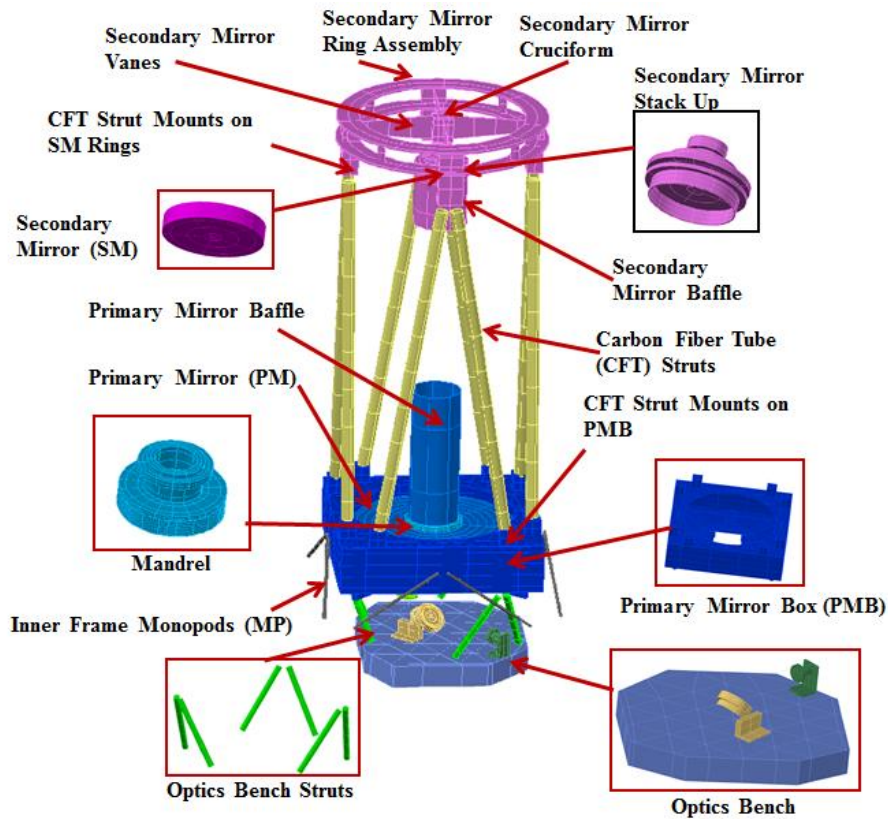

**Figure 17.** Map of telescope/bench components thermal zones into the thermal finite difference model

Figure 18 shows the predicted temperature of each structural group for the four thermal cases investigated in the STOP analysis. These structural groups are identified on the telescope in Figure 20. A number of interesting trends can be seen in these results. Firstly, as might be expected, the telescope regions near the optical bench assembly – where power-dissipating electronics and supplemental heaters are mounted – are warmer than much of the rest of the telescope assembly. The optical bench assembly is generally the warmest structural group across all thermal cases. In fact, in the hot case transient run, the optical bench is warmer than the lower limit of the thermostat dead-band, (3.5˚C), hence the heaters do not turn on. The coldest region, by far, is the carbon fiber tube telescope struts on the telescope, although it has much less dramatic change in temperature throughout the night than the secondary mirror assembly. Examining the strut mount temperatures near the secondary mirror and near the primary mirror shows the spatial gradient across these critical components, an expected result of the low thermal conductivity material of the carbon fiber struts. The middle of the strut tends to be colder than either mounting location. This trend is likely the result of the fact that the system heaters warm the primary mirror box and the secondary mirror assembly is exposed to the sun prior to the start of the observing mission phase. Carbon fiber is slow to change temperature and maintains its temperature gradients on the ascent, even through the cold atmospheric regions.

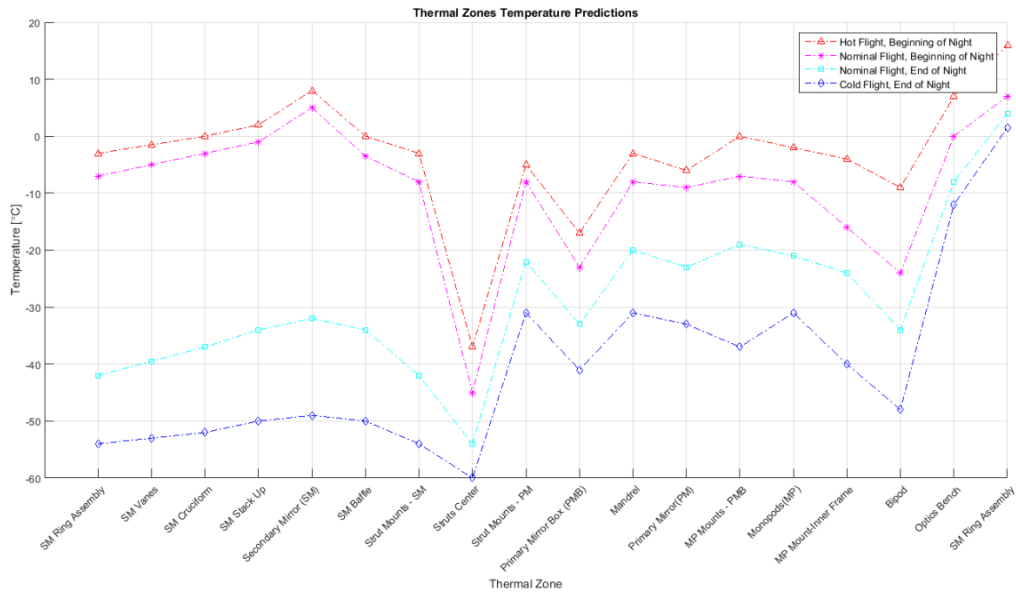

**Figure 18.** Temperatures of structural groups over all four thermal cases

It is worth noting that the WCH and WCC cases represent the hottest and coldest temperatures respectively across all of the structural groups, and the beginning of night is warmer for all structural groups than the nominal end of night. This stratification suggests that all of the structural groups follow the same temperature trends as the thermal environment is varied.

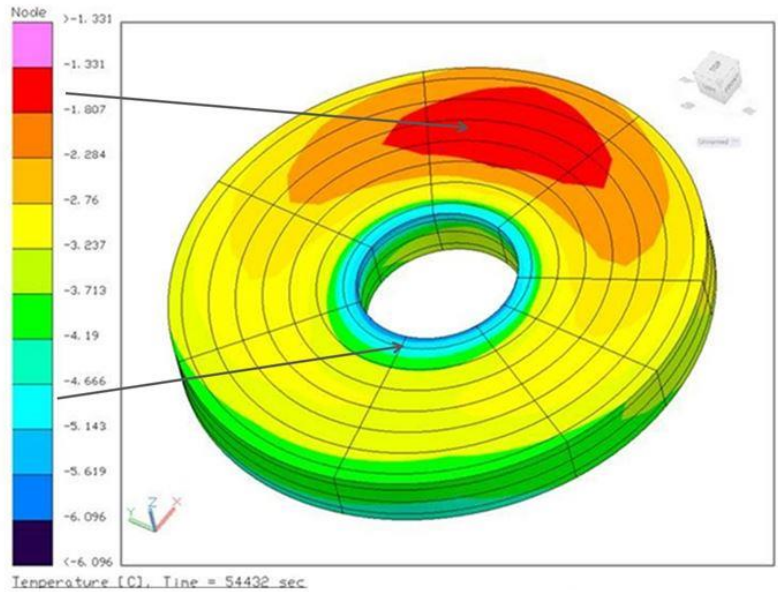

**Figure 19.** Primary Mirror temperature gradient used to assess impacts on optical performance. This gradient is a result from the worst-case hot thermal case at the beginning of nighttime and was identified as the worst in magnitude and form out of three thermal transient analysis.

The change in temperature across the telescope between the nominal beginning and end of night is much larger than the difference between the WCH and the nominal beginning of night and the WCC and the nominal end of the night. As expected, this trend suggests that the variation from one set of environmental parameters to another is less important than the significant temperature change that any individual mission will experience across a night of observing. This large temperature variation seen across missions was a major factor in the thermal design of the system and significantly influenced the telescope's structural and optical design.

In addition to temporal thermal changes, the STABLE system was evaluated to determine the spatial thermal gradients within a structural group – especially on the primary mirror. Figure 19 shows the worst-case gradient case, which occurs in the worst-case hot thermal case at the beginning of night. Because these gradients vary over time and add significant complexity in the interpretation of the STOP analysis, the results were not included in the full STOP analysis. Instead, this worst-case gradient (which was chosen because it bounded the optical performance effects of a spatial gradient) was analyzed to quantify its effect on optical parameters those results were incorporated into the optical error budget along with the temperature results for the structural groups that were used in the main STOP analysis.

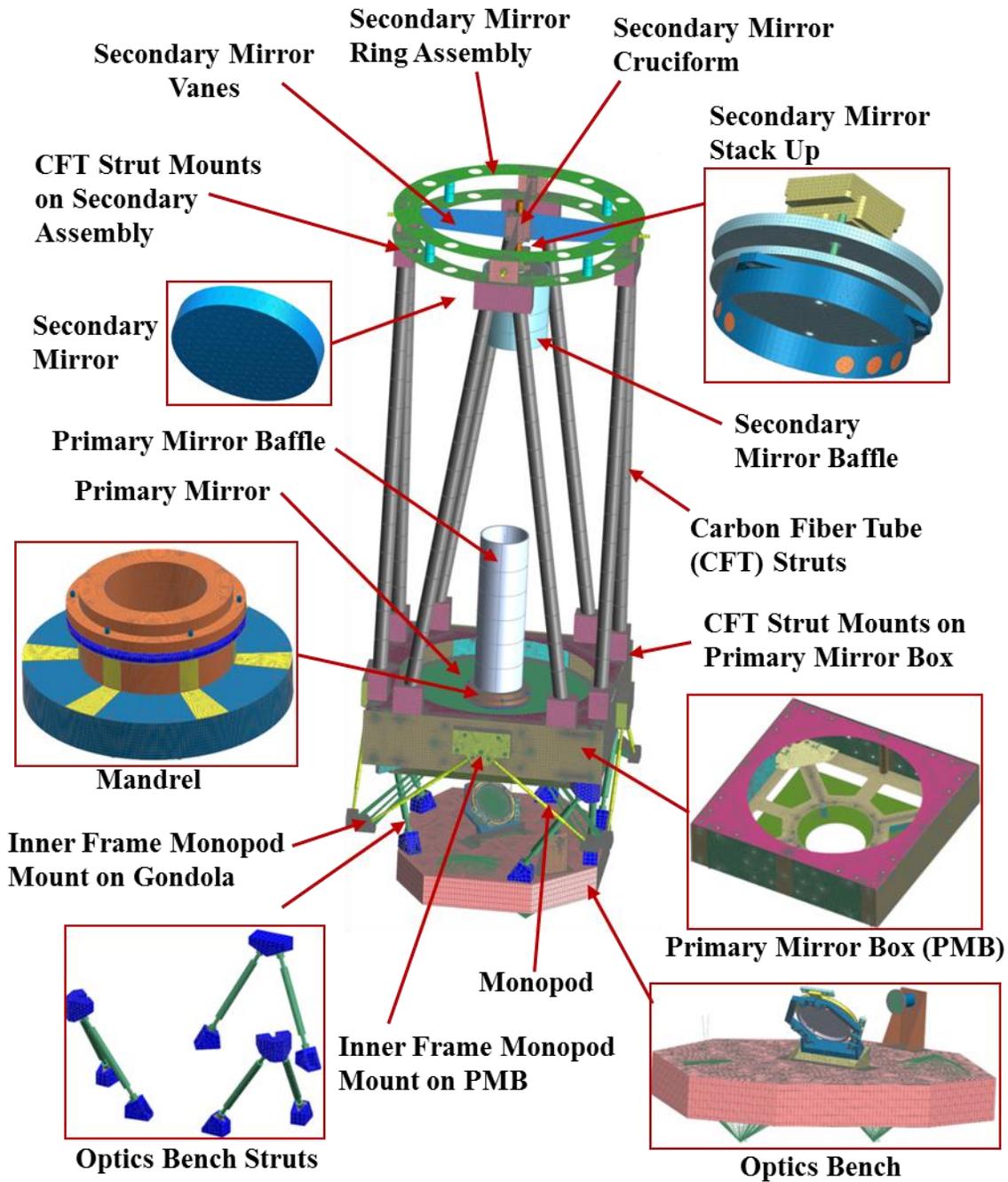

**Figure 20.** Map of telescope/bench components thermal zones into the structural finite element model

## 3  Structural Analysis

### 3.1 Structural Modeling Approach and Case Definitions

In addition to temperature, the STABLE optical performance varies over the elevation angle of the inner frame of the gondola because of different gravity conditions that induce self-deflection. The large primary mirror is not light-weighted, making the gravity influence particularly apparent in the optical performance. The gondola inner frame operates over a range of 25 to 55 degrees as measured from the horizon, so the three gravity cases considered by the STABLE STOP analysis include both of these extremes and the mean elevation angle of 40 degrees. When combined with the four thermal cases, the STABLE optical performance is evaluated over the twelve resulting cases, shown in Table 2.

**Table 2:** Case definitions based on thermal scenarios and telescope elevation angles

| Case # | Thermal Scenario | Elevation Angle |
|---|---|---|
| Case 1 | Nominal beginning of night | 40° |
| Case 2 | Nominal beginning of night | 55° |
| Case 3 | Nominal beginning of night | 25° |
| Case 4 | Worst case hot | 40° |
| Case 5 | Worst case hot | 55° |
| Case 6 | Worst case hot | 25° |
| Case 7 | Worst case cold | 40° |
| Case 8 | Worst case cold | 55° |
| Case 9 | Worst case cold | 25° |
| Case 10 | Nominal end of night | 40° |
| Case 11 | Nominal end of night | 55° |
| Case 12 | Nominal end of night | 25° |

Along with the CAD model of the system hardware, the temperature values associated with each structural group are the starting point for the structural portion of the STOP analysis. These inputs are used to develop and constrain a high-fidelity structural model, shown in Figure 20, that outputs the resulting displacement and deformation of the optical elements in the system under these thermal and gravity conditions.[11] The flow chart in Figure 21 shows the flow of the data through the NASTRAN finite element model[19] and the specialized opto-mechanical tool SigFit, in order to calculate rigid body displacements for each optical surface and their corresponding surface figure error (RMS error of the surface relative to the nominal shape). The process of this surface fitting reduces the deformation to a set of 23 Zernike coefficients.

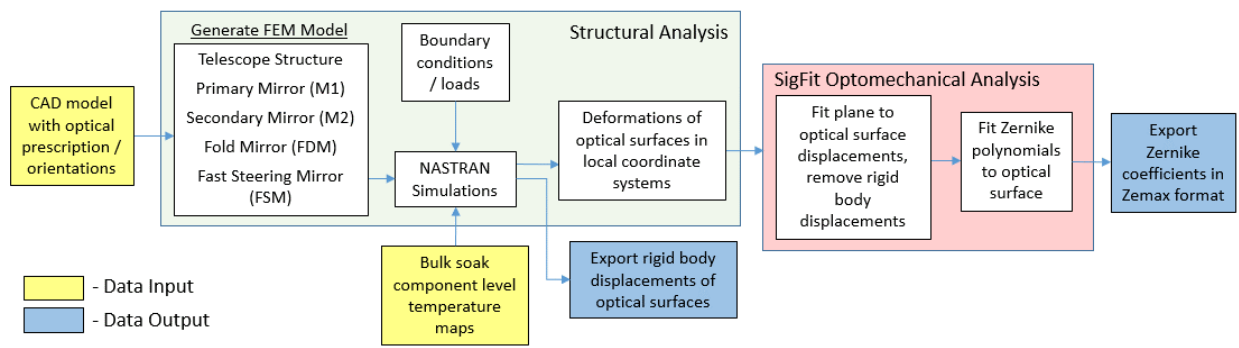

**Figure 21.** Structural analysis process flow

## 3.2 STABLE Structural Model

### Telescope Structure

A key element in assessing the optical performance of the STABLE system is the structural finite-element model of the telescope and optical bench, shown in Figure 22 with important features highlighted. Most STABLE telescope and optical bench components are modeled using thin shell and solid elements.11 The rate sensor and detector/refocusing stage assembly are modeled as lumped masses, due to the monolithic nature of their construction and lack of participation in the telescope opto-mechanics. In order to limit the total degree of freedom (DOF) count in the model, the struts are represented as beam elements with a cross-section shaped to match the geometry. Thermal gradients perpendicular to the strut axis were negligible; therefore, the implementation of higher fidelity would not contribute to the performance of the opto-mechanical model. Bolted joints are modeled with beam elements and constraint elements to join the clamped interfaces together.

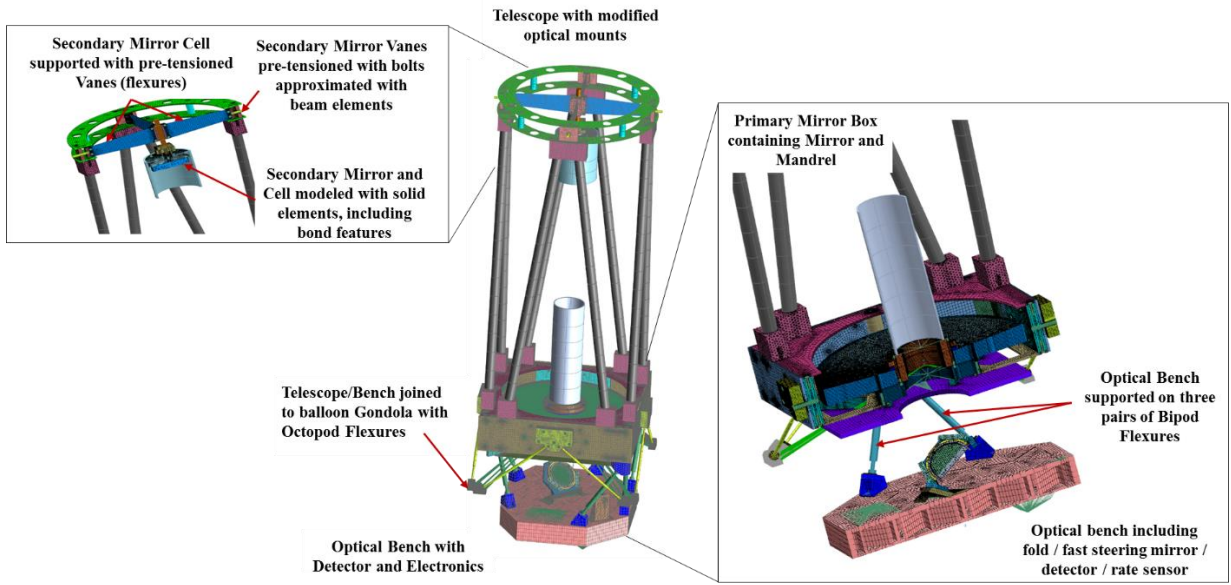

**Figure 22.** Finite Element Model features and details

The primary and secondary optical mounts on the telescope are heavily reliant on compliant pre-loaded interfaces and bolted joints, which contributes some uncertainty to its behavior under different thermal loads. The structural model attempted to compensate for these uncertainties in a number of ways: temperature-dependent material properties, detailed modeling of the structural interfaces, and high-fidelity models the primary and secondary optics and their mounts.

The performance demands of the telescope opto-mechanical model and the complexity of the optical elements drove the model fidelity to a high level for most components and interfaces. The high initial resolution of the model left few opportunities to improve model accuracy, components designated secondary to the optics judged to have little impact on the performance were modeled at low fidelity. Secondary components as the light baffles and struts did not participate in the opto-mechanical performance and were modeled with low fidelity via beam elements.

Some of the most critical elements in the finite element model are the structural interfaces, including joints, bolts, and bonded surfaces. The components of the telescope and optical bench are joined with beam elements to approximate the stiffness of bolted joints in conjunction with bonded surfaces to represent the effects of clamping. Within the instrument model, discrete bolts are represented as beam elements though the clamped areas of the interfaces associated with the optical supports are also constrained via a "glue" element to represent the joint contribution to the stiffness of the combined structure. The latter are introduced to remove some conservatism from the model predictions. Representative examples of these joint model details are shown in Figure 23. The resulting structural model demonstrated in these plots was developed using Siemens NX to automate some of the mesh generation, the level of refinement and local mesh size was determined from experience with modeling of other optical systems[19].

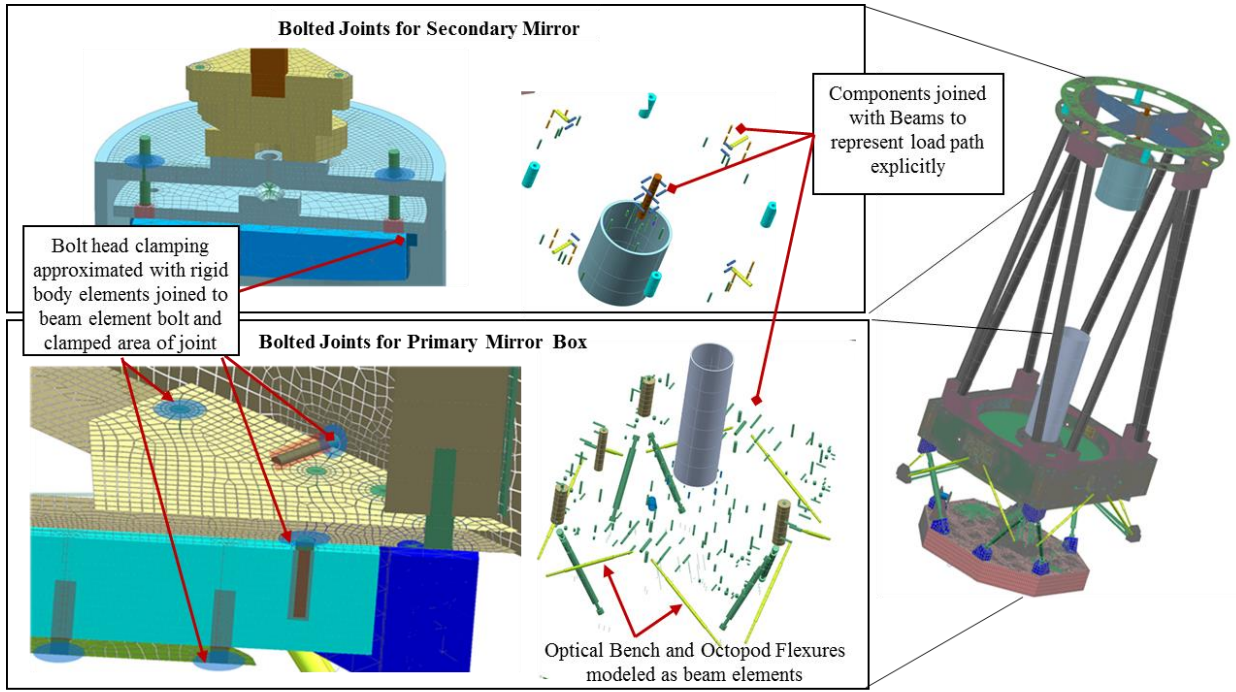

**Figure 23.** Telescope bolted joint model details; bolts and fittings modeled with beam elements

The telescope and other optical elements in the system are modeled as solid objects to capture the effect of thermal gradients and gravity loads on the final shape. The support structure surrounding each optical element is modeled with a combination of thin shell elements and solid elements, depending on the characteristics of the geometry. The primary mirror box and supporting frame for the secondary mirror are defined with thin shell elements, while the strut fittings and the cruciform supporting the secondary mirror were modeled with solid elements. Each component is modeled with an element type consistent with the topology: thin plate structures are defined with thin shell elements, while solid complex shapes are defined with solid elements.

*Primary Mirror*
The finite element models of the optical components are key to understanding the fidelity of the structural results. The primary mirror and supporting mandrel, shown in Figure 24 with a cut-away showing the interior details, are modeled with solid elements that have the aspect ratio of the mandrel internal ribs and direct contact between the mirror shim and the hub. This modeling approach ensures that the model accurately captures the "print through" of this pattern on the optical surface.

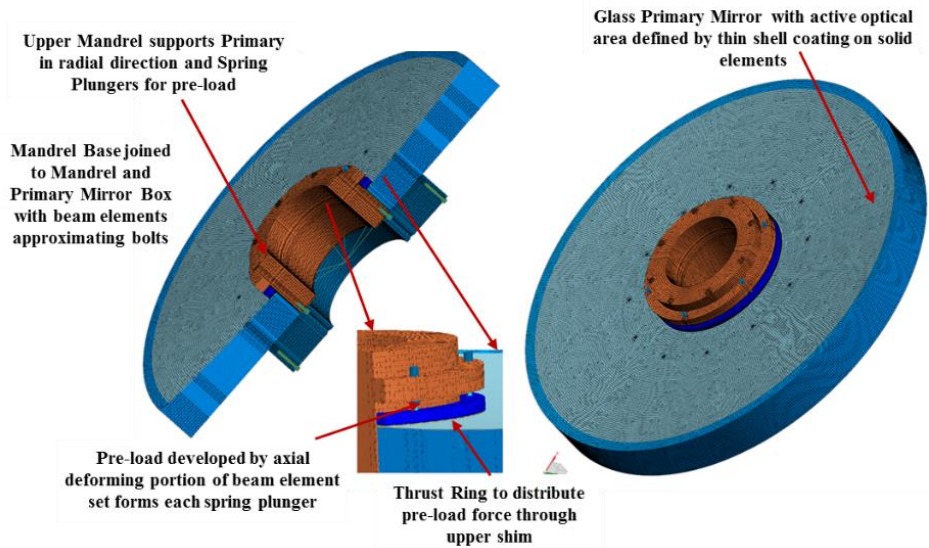

**Figure 24.** Primary mirror and mandrel finite element model

Unlike many mount designs that use a flexure system bonded to the optic, this 24 kg solid mirror is instead supported with viscoelastic shims that allow the mirror to move when subjected to gravitational and thermal loads. This floating optic/shim arrangement exhibits compliant interface behavior that required the use of a frictionless contact interface between the shim and glass. In order to ensure that the flight hardware behavior would match the modeling predictions, a layer of Teflon tape was added to the flight hardware between the shims and the primary mirror. This sliding contact improves surface figure error, but increases the rigid body motion of the primary mirror relative to the rest of the structure. The extent of the displacement depends on the elastic modulus of the viscoelastic shim material, the geometry of the shim contact, the pre-loading of the radial shims along the inner radius of the optic, and the changing gravity load due to the telescope elevation angle. In order to evaluate the optical performance at a steady-state position, the primary mirror is allowed to "settle" into its new equilibrium position prior to calculating deformation for the optical figures of merit. This technique is implemented by evaluating the first pre-load/gravity load case twice: the first load case to establish mirror boundary conditions, and the second to ensure the primary is in equilibrium prior to evaluating different thermal/gravity orientations.

The spring plungers that maintain a preload on the primary mirror in the vertical (through the thickness) direction are another key element in the primary mirror mounting structure because they maintain the position of the mirror along the mandrel/piston degree of freedom. As shown in Figure 24, the spring plungers apply a preload through a thrust ring that compresses the viscoelastic shim that rests on the active surface of the mirror. The pre-loading is implemented by axially deforming the plunger body to a desired length and allowing the upper shim to compress. In the NASTRAN model, these spring plungers are represented as beam elements with a spring element between the Delrin tip and the steel plunger / bolt. In the model, the plunger tips are connected to the thrust ring via stiff axial springs and soft lateral springs to prevent excessive lateral motion of the upper shim during telescope elevation angle changes.

*Secondary Mirror*

The secondary mirror is another important element in the STABLE optical system. The STOP analysis provided insight into the mount design, resulting in a flexure ring that forms the mirror cell which bonds to the optic circumferentially. In the model, the secondary mirror support is defined with solid elements for the bond lines and thin shell elements for the flexure ring that forms the mirror cell, as shown in Figure 25. As with the other large optical elements in the telescope model, the solid elements provide a surface for evaluation of optical figures of merit and can be deformed with gradient loads. The bolted connections to the rest of the telescope are represented with beam elements, consistent with the rest of the telescope model.

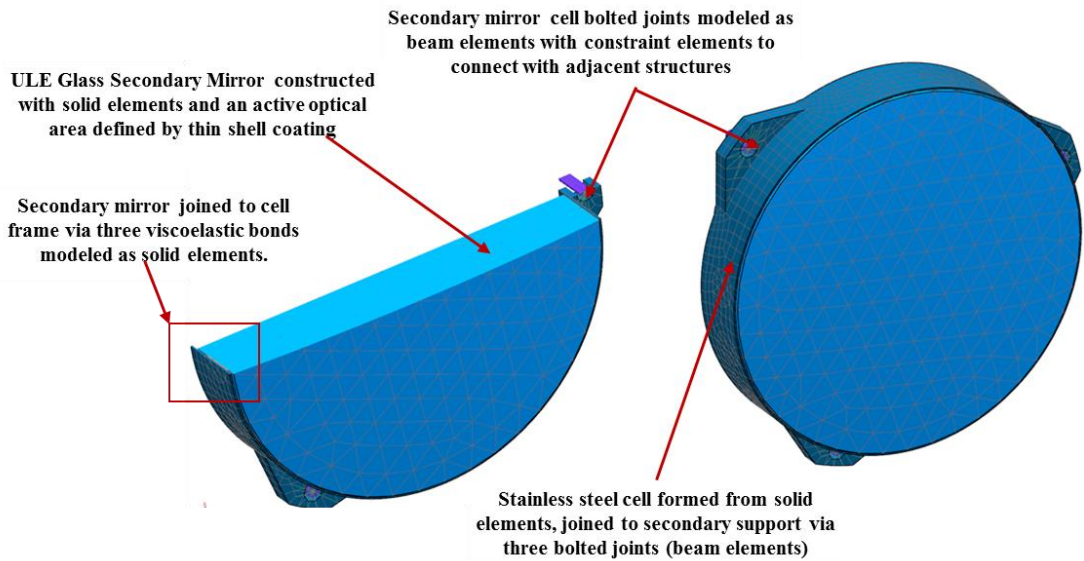

**Figure 25.** Secondary mirror structural modeling details

*Optical Bench with Fold Mirror and Fast Steering Mirror*

The optical bench assembly, which includes the optics and electronic components mounted to the optical bench, follows a modeling philosophy consistent with the telescope modeling approach. The thick bench structure was modeled explicitly using solid elements, along with the fast steering mirror and fold mirror. The supporting brackets for the large bench optics are modeled with thin elements, consistent with topology of those components. Beam elements are used to represent the hexapod struts, and flexures supporting the fold mirror for similar reasons noted in prior components. The electronics and detector stage are modeled with a rigid mass joined to the optical bench, to limit the complexity of the structural model and focus the modeling efforts on errors affecting the optical path. Performance uncertainties were included in the optical model of the system to compensate for this lack of fidelity. The bench model provides a means to check requirements and design margins separate from the telescope, but it also provides an optical-mechanical representation of the fold mirror, fast steering mirror, detector, and supporting structure, rounding out the end of the optical path to enable a performance analysis of the whole instrument.

*3.3 Structural Outputs*

The STABLE structural models were used to solve a variety of design and analysis challenges. In some cases, the analyses were performed using component models rather than the full payload structural model. The results were then included in the system error budget roll-ups. For example, the full STABLE STOP analysis implemented a single temperature for each component including the optics to evaluate optical performance over the flight. However, to capture the performance degradation due to spatial thermal gradients, a worst-case gradient (determined to be beginning of night) was mapped onto the primary mirror standalone model and used to characterize the mirror's optical performance when subjected to the gradient. The Zernickes produced were then used to perturb the optical model and develop error terms as a function of this gradient, which were ultimately included in the optical error budget. Figure 26 shows how each component model was used to determine various elements of the error budget or answer specific design questions. All of these components were included at some level in the full STABLE structural model that was used to perform the complete opto-mechanical assessments on the system. This work addresses the optical performance of the system and not the specific design decisions or ancillary outputs produced by the modeling effort.

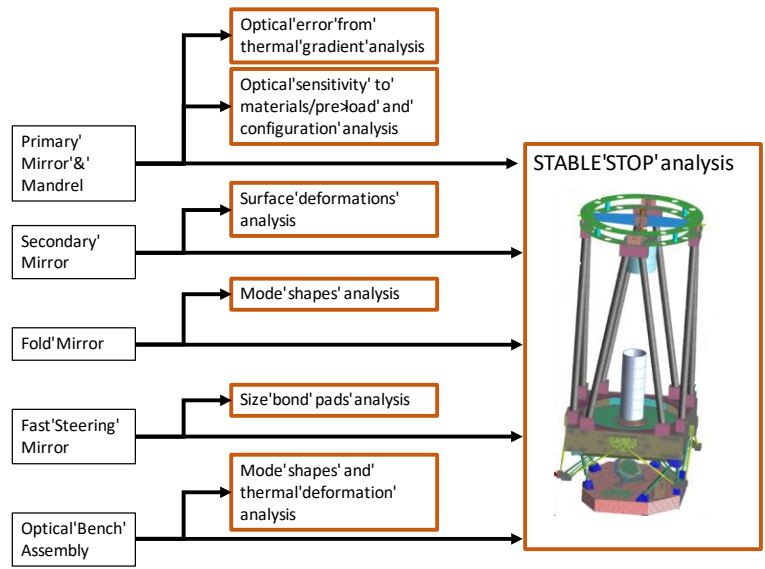

**Figure 26.** Process Flow from Structural Model Analysis Products

The optical surface deformations and rigid body displacements of the optical elements are calculated for each of the 12 loading scenarios indicated in Table 2, as well as a baseline ground-aligned scenario. The different environments are computed using the telescope/bench thermal model described in Section 2, and mapped to the structural model described in Section 3.2 STABLE Structural Model. The optical displacements developed by the ground calibration case are subtracted from the subsequent flight cases to emulate the process of optical alignment prior to the flight. The optical surface displacements of the structural model are reduced to Zernike terms via SigFit.[20]

The STABLE STOP analysis then uses the SigFit software to calculate optical performance for a system of optics due to mechanical effects. The software converts surface deformations of a given optic to the Zernike representation of the deformed shape, which can then be used to evaluate the performance of the distorted optical system.[20]

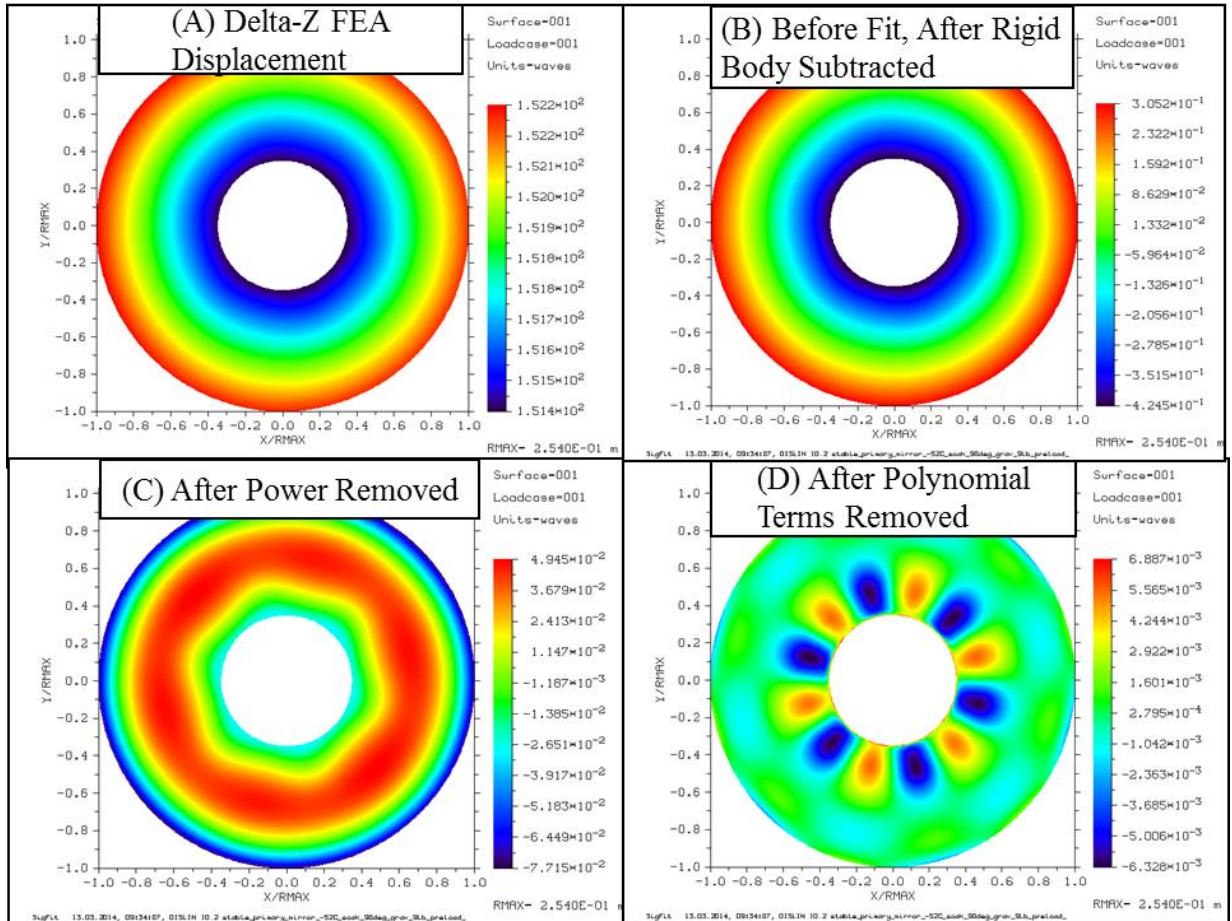

**Figure 27.** Example of SigFit deformed surface fitting process

The structural results in the STOP analysis are often intermediate products that are not directly interpreted until they are processed through the optical analysis in order to determine wavefront error and other performance metrics. However, Figure 27 and Table 3 show two examples of the structural outputs for a single case: a 40 N pre-load applied to the mirror top surface and -52C cold soak of the mirror/mandrel. Figure 27a shows the vertical displacement of the optical face with respect to the local coordinate frame (positive into the page), as generated from the structural analysis. The rigid-body displacements of the optic are included as a separate input to the optical analysis, in the form of a table like those shown in Table 3. Those displacements are therefore removed during the SigFit deformed surface fitting process so the flexible response of the optic remains, as shown Figure 27b. The second two plots show the effect of removing the main terms derived from the fit process: power and polynomial terms. The power term shows the change in focus of the optic and can therefore be removed by refocusing the system. Figure 27c has the power term removed and shows a perfectly-focused optic with the effect of the mandrel clamp and boundary conditions on the primary. Figure 27d demonstrates the residual deformation after removal of the polynomial terms of the Zernike and therefore reveals the higher order shape of the primary, driven by the position of the shims and the preloaded spring plungers. Although the last two plots generated during the optical fit process are not used directly in the optical model, the information contained in the plots provides useful insight into the mechanical behavior of the optical mount. The Zernike terms are calculated after tip/tilt/piston is removed, shown in Figure 27b, and exported as a table for integration with the optical model of the telescope-IOBA system.

Table 3. Example of rigid body motions output from the structural analyst

| | Rigid body motions | | | | | |
|---|---|---|---|---|---|---|
| | dX [m] | dY [m] | dZ [m] | ΘX [rad] | ΘY [rad] | ΘZ [rad] |
| Primary | -1.723E-05 | -1.819E-04 | -3.738E-05 | -9.560E-05 | 1.970E-06 | 2.023E-05 |
| Secondary | -6.986E-07 | -2.376E-04 | -1.557E-05 | 2.474E-04 | -4.057E-06 | -1.020E-05 |
| Fold | -6.521E-05 | -2.029E-04 | -8.460E-05 | 1.826E-05 | -4.567E-05 | 3.279E-05 |
| FSM | -9.695E-05 | -1.951E-04 | 1.295E-04 | 8.367E-05 | 8.068E-05 | 3.109E-05 |
| Detector | -1.219E-04 | -2.021E-04 | -2.444E-05 | 1.826E-05 | 9.419E-08 | 2.023E-05 |

## 4    Optical Analysis

### 4.1 Optical Prescription Modeling Approach

The final step of the STOP process, the optical modeling, uses the two data sets generated by the SigFit tool described in the previous section (the optical surface deformations as represented by the Zernike coefficients and the rigid body motions) and combines them with the surface figure fabrication errors. Starting from a perfectly aligned and perfectly fabricated optical prescription, these errors are then incorporated into the Zemax prescription model. From here, the system is refocused to remove errors that would be addressed by the STABLE refocusing system. Similarly, the analysis also removes errors associated with the decenter of the wavefront to capture the fact that STABLE is aligned prior to launch. The resulting wave front error from the optical model is then combined with other bounded errors not captured in the twelve STOP analysis cases via the system optical error budget. This error budget includes error terms such as the expected calibration and alignment residual errors, refocusing residual errors, and worst-case errors due to thermal gradients. This final value then provides the estimate of the total system performance across all twelve thermo-mechanical cases, which in turn, is used to inform design decisions and understand sensitivities in the system.

### 4.1.1 Surface Figure Fabrication Errors

A modeling technique is used to simulate the maximum allowable surface figure fabrication errors in all four STABLE mirrors. This technique utilizes Power Spectral Density (PSD) modelling to represent mid- and high-spatial frequency components of the optical surface height errors, as described by Sidick.[21] The magnitude and spatial frequency of these errors represents the wavefront error (WFE) associated with typical mirror figuring processes.

A MATLAB script generates a dataset of fabrication distortions for each of the mirrors, where the magnitude of these distortions depends on the diameter and expected fabrication errors of each mirror. The telescope vendor was required to deliver a telescope with a WFE no larger than 0.114λ RMS at 633nm. The primary (M1) and secondary (M2) mirrors are thus modeled to split the WFE contribution evenly: 0.057λ RMS at 633nm for each. The commercial fold and fast steering mirrors both have specifications that list the quality of the mirror as λ/20 RMS at 633nm. The resulting surface deformations for each mirror are shown in Figure 28. These four surface fabrication error datasets are then incorporated into the optical prescription as Grid Sag surfaces. For reference, the total system wavefront error including these fabrication errors but excluding misalignments, thermal effects, preloading of M1, and gravity sag, is .0346λ RMS at 633nm. This corresponds to a Strehl ratio of .854.

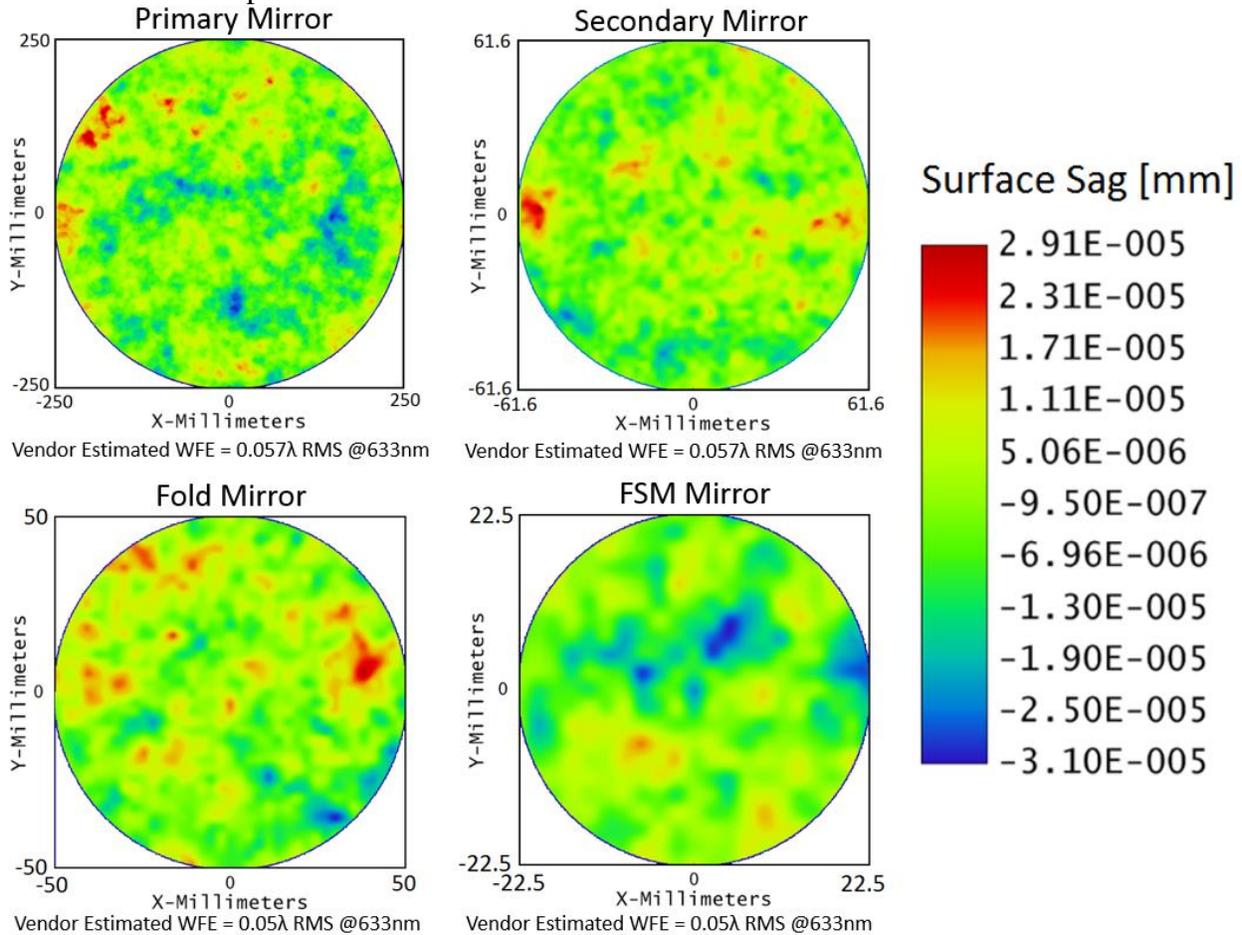

**Figure 28.** Surface figure irregularities for each mirror. These represent the manufacturing errors that each surface contributes to the optical performance of the system.

## 4.2   Optical Performance Results

### 4.2.1 Structural Model Validation and Trade Studies

#### Primary Mirror Model Validation

The primary mirror mount is undoubtedly the most complex assembly to model in the STABLE system, and was the target of many modifications to minimize indeterminate mechanical effects and improve performance against external disturbances. The results from this model were validated by performing a standalone study analyzing the effects of individual loads on the mirror shape. The study examined four external applied loads: mirror temperature, spring plunger preload, telescope elevation angle, and gravity sag effects. Each was evaluated separately so their contributions to system WFE are seen directly These results are shown in Figure 29.

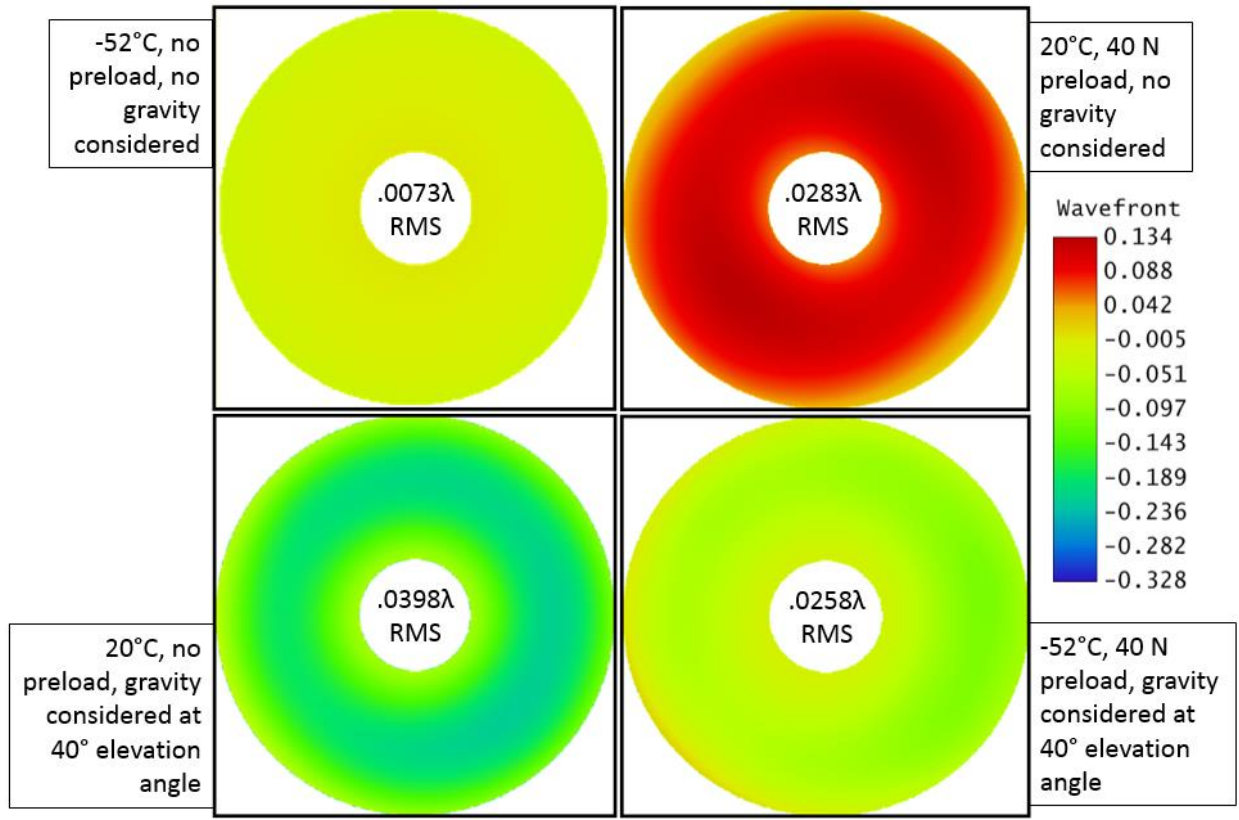

**Figure 29.** Primary Mirror wavefront error contributions evaluated separately. Red indicates a negative axial deflection (into the page) and blue indicates a positive axial deflection (out of the page). The units are in λ at 633nm.

Starting from the top left, it is clear that the effect of a exposing the primary mirror to a low temperature has a minimal effect on the WFE. Because the mirror is made of class-0 Zerodur, which has a very low coefficient of thermal expansion, it is not expected to change shape significantly over a wide range of temperatures. The next figure in the top right shows the effect of adding a 40 N preload from each spring plunger. The 40 N preload/plunger was an initial estimate of the value needed on the system; STABLE's hardware actually uses 15.6 N of preload in each of the six spring plungers (93.6 N preload total). The preload of the spring plungers does generate a significant wavefront error, at 0.0283λ RMS at 633nm. The preload from the spring plungers squeezes the center of the mirror, as shown by the ring of red around the center bore. The figure in the bottom left shows the gravity sag in the nominal 40° elevation case, as measured from when the optical axis of the telescope is parallel with the ground. As can be seen in the image, the mirror sags about the center mandrel, which makes the mirror droop around its mount like a mushroom. However, as shown in the bottom right figure, when all of these forces are combined, the preload and cold soak counteract the drooping generated by the gravity sag. Thus, the overall WFE on the primary mirror is smaller than in the gravity sag case alone, at 0.0258λ RMS at 633nm.

These initial results provide critical insight into the behavior of the overall system behavior. Gravity sag is clearly the dominant source of error on the primary mirror, although the error due to the preload of the spring plungers is also a significant contribution. The design decisions made to reduce the thermal contraction of the mirror clearly limit the impact of a thermal cold soak on the WFE, but because this effect counteracts the mirror deformation due to the gravity sag, the analysis has more complex trends that are explored in Section 4.2.2 Optical Performance Metrics.

*Thermal Gradients on the Primary Mirror*

Thermal gradients are another potential contributing factor to the system's performance degradation, but the STOP analysis approach used on STABLE uses bulk soak temperatures for all components, including the primary mirror. In order to determine if this assumption was valid, a breakout finite element model study was performed to estimate the optical performance degradation due to thermal gradients on the primary mirror (which, as the largest optic in the system, is most susceptible to larger thermal gradients).

**Table 4.** Effect of Thermal Gradients on Optical Performance When Modeling the Primary Mirror

|  | Case 1 | | Case 2 | | Case 3 | |
|---|---|---|---|---|---|---|
|  | No Gradient | Gradient | No Gradient | Gradient | No Gradient | Gradient |
| Strehl Ratio | 0.616 | 0.622 | 0.484 | 0.485 | 0.734 | 0.740 |
| Spot Decenter [mm] | 0.844 | 0.844 | 0.770 | 0.770 | 0.768 | 0.768 |
| WFE [waves RMS] | 0.1108 | 0.1097 | 0.1356 | 0.1354 | 0.0885 | 0.0873 |
| Focus Shift [mm] | 2.875 | 2.875 | 3.245 | 3.245 | 2.445 | 2.445 |

Table 4 details the results of this study, although they do not represent the final performance of the system as the analysis was performed at an early phase and used solely to compare bulk soak and gradient temperatures to bound the likely effect of gradients on the system. Clearly, the effect of thermal gradients on the primary mirror is relatively small (on the order of $0.002\lambda$ RMS at 633nm). In fact, optical performance improved slightly when thermal gradients are included, likely due to a similar effect observed in the primary mirror model validation study: the thermal deformation of the system counteracts the gravity sag deformation. This breakout study was performed only for the first 3 cases as a means to verify that the thermal gradient effect on the primary was small. As such, it represents a best guess nominal thermal gradient and not all thermal gradients that may be found in the primary. For simplicity, the STABLE STOP analysis reports results for bulk soak temperatures for all components and including the primary mirror, and includes an allocation for thermal gradient effects in the system performance error budget.

*Primary Mirror Mount Spring Plunger Preload Study*

As seen in the primary mirror model validation study, the spring plunger preload is a significant factor in the system's WFE. Therefore, STABLE performed a design study to determine the optimum spring plunger preload to both secure the primary mirror in place on the mandrel and to minimize distortions of the optical surface. This spring plunger interacts with a viscoelastic shim as shown in Figure 24. Although other studies were performed to determine the appropriate material for the shims, this spring plunger study used the material Nusil CV2-2566 (which has a constant storage modulus between room temperature and -100°C ) because it was used in building the actual STABLE hardware. A sample of the spring plunger preload assessment results is shown in Figure 30. The point spread function of the system degrades with larger preloads, but the smaller preloads carry a higher risk of primary mirror motion along the mandrel, which is difficult to model. Ultimately, STABLE chose to balance this risk by selecting a preload value of 15.6 N per plunger (for a total of 93.6 N of total preload applied by the six spring plungers). This value meets the 0.6 Strehl Ratio requirement with a small margin. All subsequent analysis used these primary mirror mount design parameters.

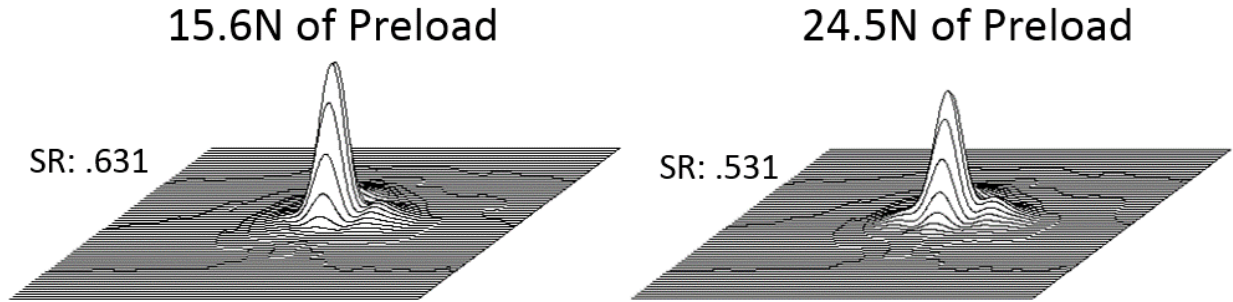

**Figure 30.** PSFs and Strehl ratios showing the effect of spring plunger preload on the primary mirror. These results are generated using the Case 1 scenario (nominal beginning of night thermal scenario with a 40° telescope elevation angle).

### 4.2.2 Optical Performance Metrics

The resulting system optical performance at the twelve flight cases (shown in Table 2) is characterized using a number of metrics. For STABLE, the most important optical performance metrics are:

1)  Total system wavefront error and Strehl ratio, which summarize the image quality of the target star on the detector and help the pointing analysis team evaluate the effectiveness of their stabilization routines
2)  Spot decenter on detector, which is used to ensure that the detector's field of view is large enough to capture the image of a target star as it shifts due to these errors
3)  System focus shift, which is used to ensure that the range of the refocusing stage is sufficient to center the detector on the focus of the optical system
4)  Change in effective focal length, which effects the computation of the system pointing stability metrics, and
5)  RMS spot radius, which affects the system signal-to-noise, windowing processes in the pointing control software, and is related to the system's image quality and pointing stability

The twelve flight cases were then used to bound the system behavior, identify trends, and highlight sensitivities of the parameters over a range of conditions.

### Total Thermally- and Mechanically-Induced Wavefront Error

As shown in Figure 31, the total WFE generated by these twelve cases helps to determine how the various temperature scenarios and elevation angles impact the optical performance. As was expected, the colder the environment, the larger the WFE. This effect is driven by thermally-induced misalignments and optical surface deformation from each mirror's mount. These results show that increases in elevation angle correspond to larger WFE. The main driver of this effect on WFE is the gravity sag on the primary mirror, as shown in the primary mirror model validation. The telescope structure itself actually sags less at high elevation angles. Overall, the worst WFE across elevation angles is expected at the worst-case cold temperature profile, whereas the worst-case hot temperature profile tends to produce the best WFE. The change in WFE over the nominal temperature profile from the beginning of observing to the end of observing mirrors the thermal changes observed in the thermal results in Figure 18. It is worth noting, however, that elevation angle changes (which can occur in between observations approximately every 10 minutes as STABLE moves to a new target star) can cause significantly larger changes in wavefront quality compared to this temperature change throughout the night. One possible solution to limit this effect is to limit the elevation angles of target stars.

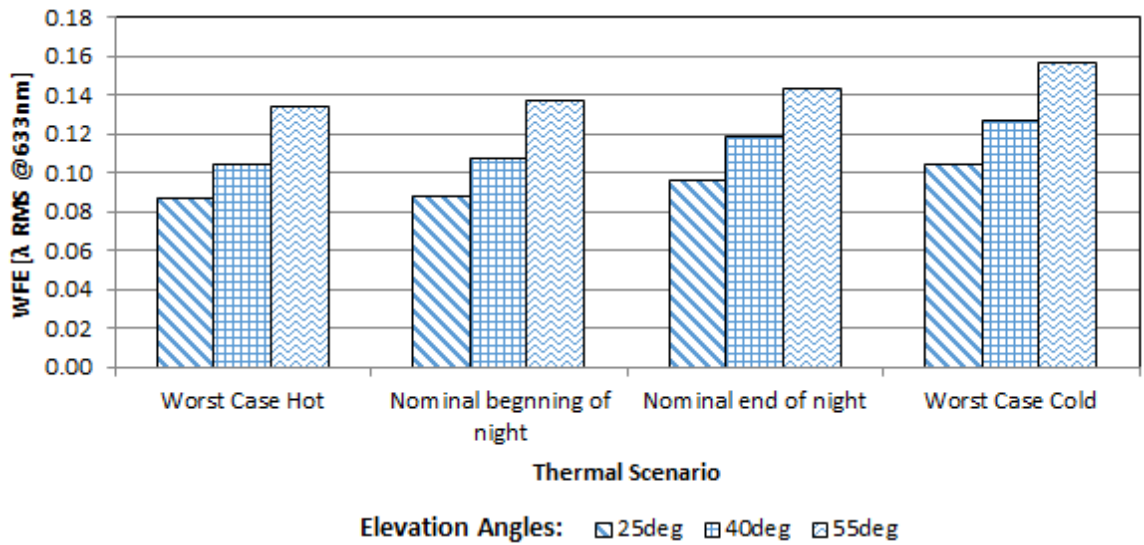

**Figure 31.** Wavefront error at various temperature and elevation angle scenarios

*Strehl Ratio*

Figure 32 shows the Point Spread Functions (PSF) and Strehl ratio for each of the twelve cases. As the Strehl ratio is related to the system WFE, the same trends apply, but this representation enables a better evaluation of the spot quality predicted for different conditions. A Strehl ratio of 0.6 (or higher) was required of the system in order to ensure it could achieve the 100 milliarcsecond over 60 second (1-σ) pointing stability, specifically in the nominal beginning of night case. It is clear from the figure that this requirement is met, and that that the system can achieve acceptable optical performance at lower elevation angles and warmer temperatures. However, the system cannot achieve this performance for the highest elevation angle, and during a nominal mission may not meet the requirement by the end of the night at the nominal elevation angle. This trend suggests that higher-elevation targets are best suited to the beginning of the observation period, and optical performance can be improved by finding lower-elevation targets.

| | Worst Case Hot | Nominal Beginning of Night | Nominal End of Night | Worst Case Cold |
|---|---|---|---|---|
| 55° Elevation Angle | 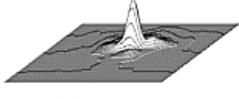 SR = .492 | 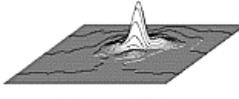 SR = .474 | 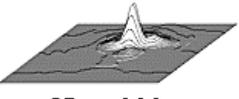 SR = .444 | 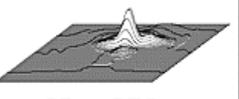 SR = .376 |
| 40° Elevation Angle | 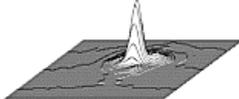 SR = .650 | 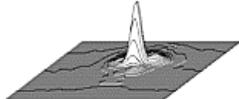 SR = .631 | 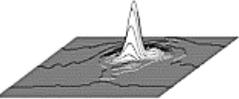 SR = .573 | 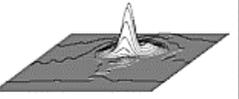 SR = .528 |
| 25° Elevation Angle | 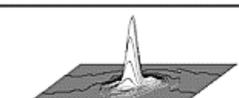 SR = .739 | 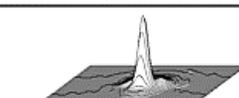 SR = .733 | 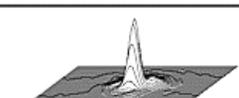 SR = .694 | 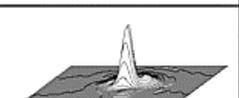 SR = .647 |

**Figure 32.** Point Spread Functions and Strehl ratios at various temperature and elevation angle scenarios

*Spot Decenter on Detector*

Because the accuracy of the pointing is not a STABLE requirement (the stability is the main focus of the demonstration), the spot decenter is tracked as a system metric primarily to ensure that the guide mirror throw is enough to support the predicted decentering motion throughout a variety of conditions. STABLE's error budget allocation for maximum spot decenter is 2.2mm. This metric also enables the pointing control algorithms to be properly tuned to allow for this shift in position. The results, shown in Figure 33, shows that the decenter follows the same temperature trend as WFE: lower environmental temperatures correspond to larger spot decenter. Clearly, the temperature conditions have a larger influence than the elevation angle on this particular parameter. The elevation angle trend – with the nominal angle generating slightly more decenter than either of the two elevation extremes – is a result of how the primary mirror and the rest of the telescope structure are deformed under gravity sag. As the elevation angle increases, the primary mirror sees a larger tilt effect due to gravity sag. The structure on the other hand will tilt less as that angle increases.

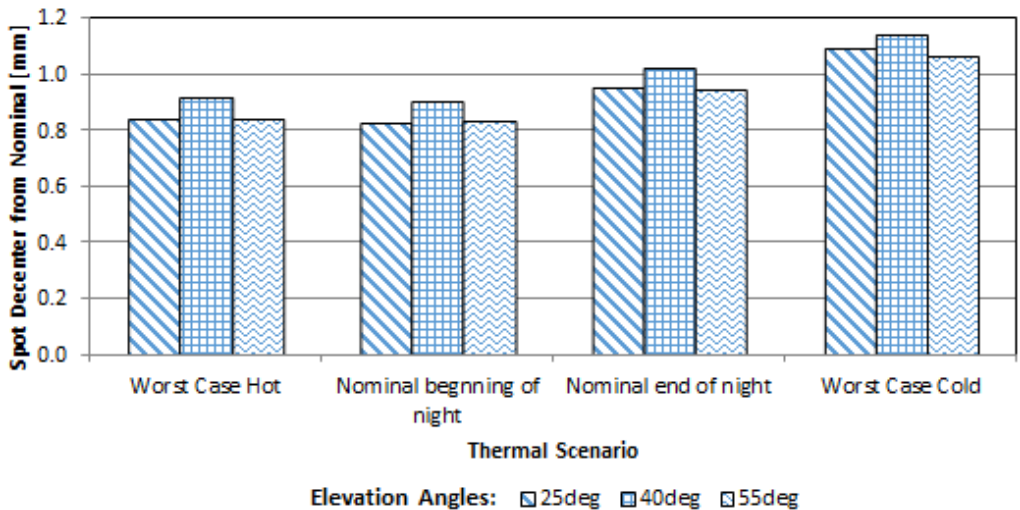

**Figure 33.** Spot decenter on detector at various temperature and elevation angle scenarios

*System Focus Shift*

The system focus shift represents the amount that the system focus moves from the ground-aligned focus position. Because STABLE has a refocusing stage, and those bounding errors are included in the system's optical error budget rather than in the STOP analysis, this metric is primarily used to ensure that the refocusing stage has enough stroke to position the stage at best focus in all of the STABLE mission use cases. Shown in Figure 34, the system focus shift has a few notable trends. As with decenter, the thermal conditions tend to dominate the focus shifts, with elevation angle contributing relatively small variations in focus at the same thermal conditions. In every case, the focus shifts further away from the telescope because the thermal contraction of the telescope decreases the primary mirror and secondary mirror spacing, which pushes the system focus further out. At the same time, the cold optical bench contracts, which brings the optical elements closer together. This, too, pushes the system focus further out. The combination of these effects results in the plot below, where the nominal case at the beginning of the observation generates higher focus shifts than the worst case cold or worst case hot conditions. The magnitudes of these shifts rely upon the temperatures at specific components along the optical path. Each thermal scenario provides a different combination of component temperatures; if a simple bulk soak temperature is used for the whole telescope, we would expect to see a linear trend in how the focus shifts. The sensitivity of these results to the specifics of the component temperature leads to questions about whether these twelve cases have properly bounded the potential change in focus location. In order to address this problem, the system was designed with significant margin in the translation stage stroke. The largest change in focus in the twelve cases was approximately 4 mm. The STABLE translation stage has 50 mm in stroke, and the telescope is aligned to center the ground-aligned focal plane at the center of the range of motion, providing 25 mm of stroke in the focus shift direction indicated by the analyses, so the stage clearly maintains enough stroke to meet STABLE's needs.

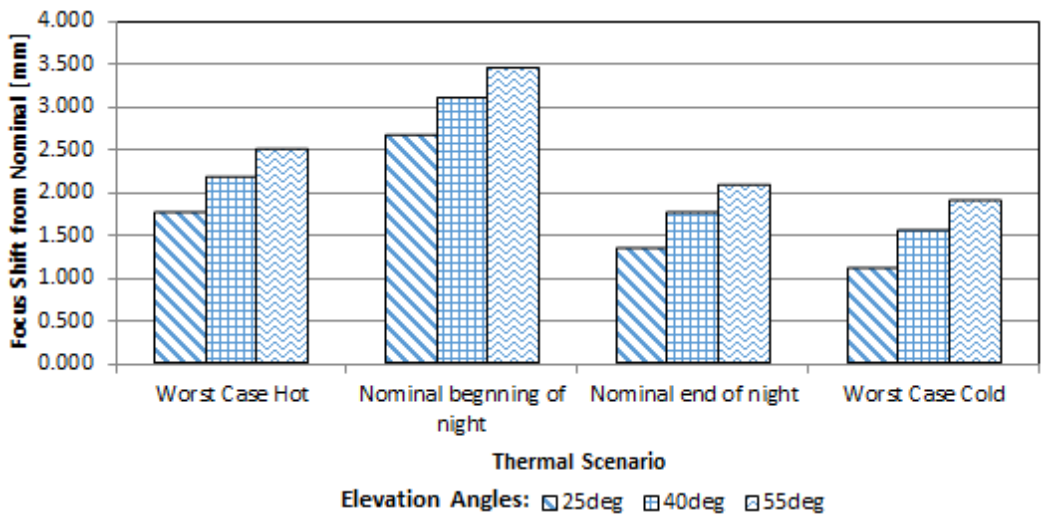

**Figure 34.** System focus shift at various temperature and elevation angle scenarios

*Effective Focal Length*

The effective focal length (EFL), shown in Figure 35, is a critical parameter that is used for mapping the system pointing performance on the sky and computing parameters in the pointing control algorithms. This parameter is primarily governed by the behavior of the primary and secondary mirror and, in general, it decreases from the ideal value of 9000mm. Because STABLE uses this parameter to compute the pointing performance of the system, the total system performance assessment is dependent on the knowledge of this parameter. The BIT-STABLE has no direct measurement of the EFL, so it imposed a requirement that the parameter not change by more than 4.5 mm in order to ensure that the pointing stability requirements are met even if the EFL deviates by this much in flight. However, the figure shows that this metric follows an irregular temperature trend: at the nominal temperature case at the beginning of observing shows a smaller decrease than the other temperature scenarios. As with the system focus shift, the magnitudes of the EFL changes rely upon the temperatures at specific components along the optical path. Each thermal scenario provides a different combination of component temperatures. As with the system focus shift, if a simple bulk soak temperature is used for the whole telescope, we would expect to see a linear trend in how the EFL changes. The elevation angle trend is likely a result of how the primary mirror and the rest of the telescope structure sags with gravity. As the elevation angle increases, the primary mirror sees a larger surface sag effect due to gravity. This surface sag effect can change the conic value and radius of curvature, which in turn affects the EFL. The telescope structure deformation also plays a role, as any change in the primary mirror-secondary mirror spacing has a large impact on EFL. Clearly, the EFL change requirements can be met in warm cases and at the nominal beginning of night case, but as the system cools, this parameter exceeds the requirements.

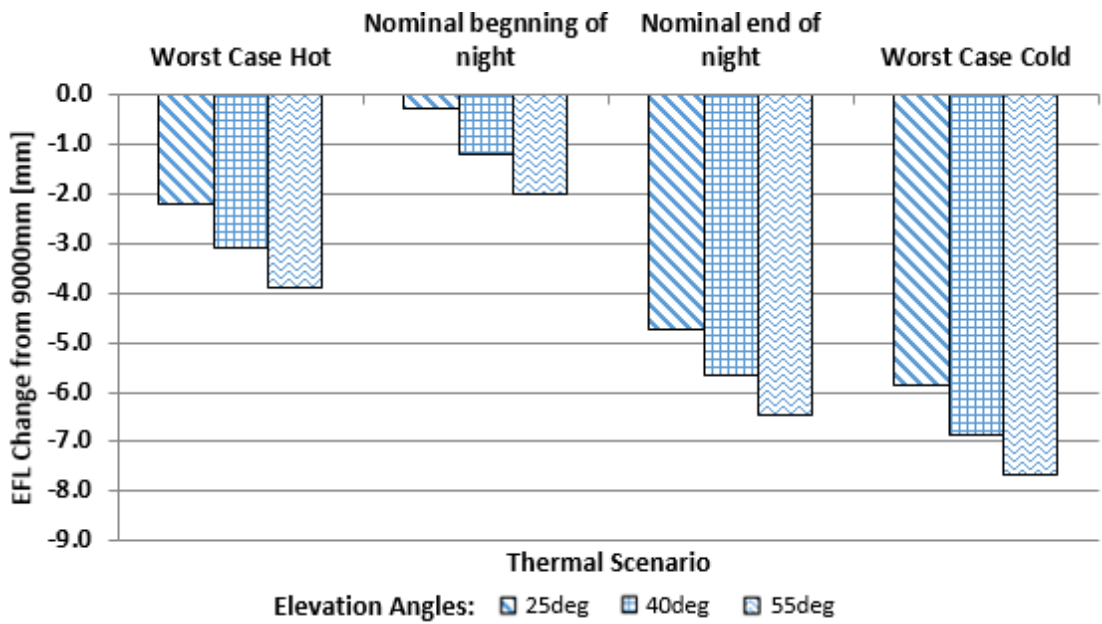

**Figure 35.** Effective focal length change from 9000mm at various temperature and elevation angle scenarios

*Spot Radius*

The spot radius, shown in Figure 36, is also an important parameter because it influences the minimum window size that the software can choose to collect in each frame of the camera. This parameter also follows an irregular temperature trend: the rigid body motions and surface deformations couple together in ways that can influence the spot size. The elevation angle trend is likely a result of how the primary mirror and the rest of the telescope structure deform under gravity sag. As the elevation angle increases, the primary mirror sees a larger tilt effect due to gravity sag. The resulting tilted spot appears larger than a spot without added tilt.

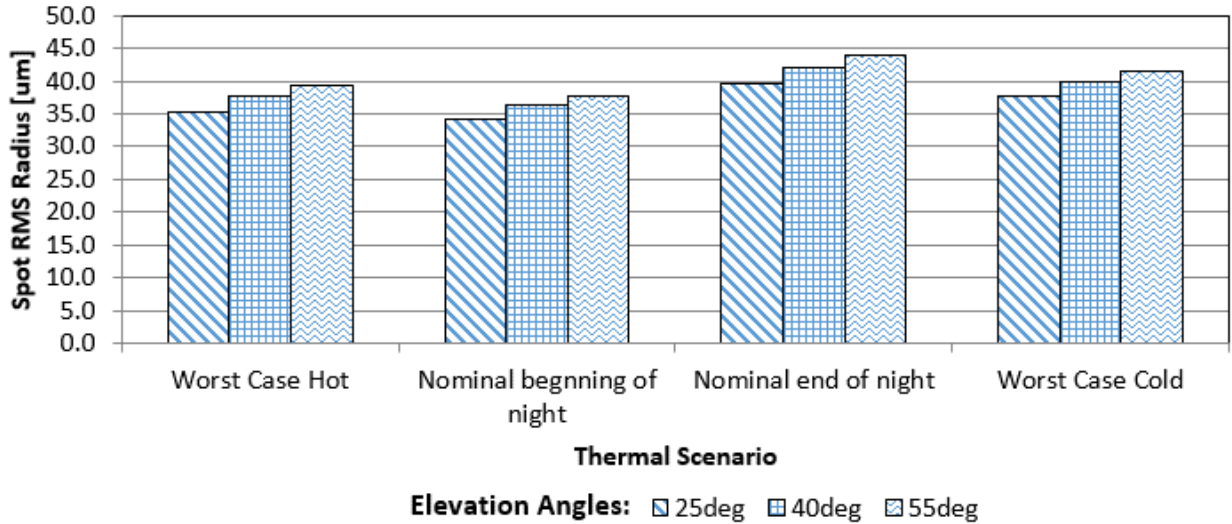

**Figure 36.** Spot RMS Radius at various temperature and elevation angle scenarios

### 4.2.3 Optical Performance Sensitivities

The four optical performance metrics used to understand STABLE's sensitivity to telescope elevation angle and thermal scenarios are wavefront error, spot decenter, system focus shift, and spot radius. Figure 37 and Figure 38 show the impact on these performance metrics when the telescope elevation angle changes and when the thermal scenario changes, respectively.

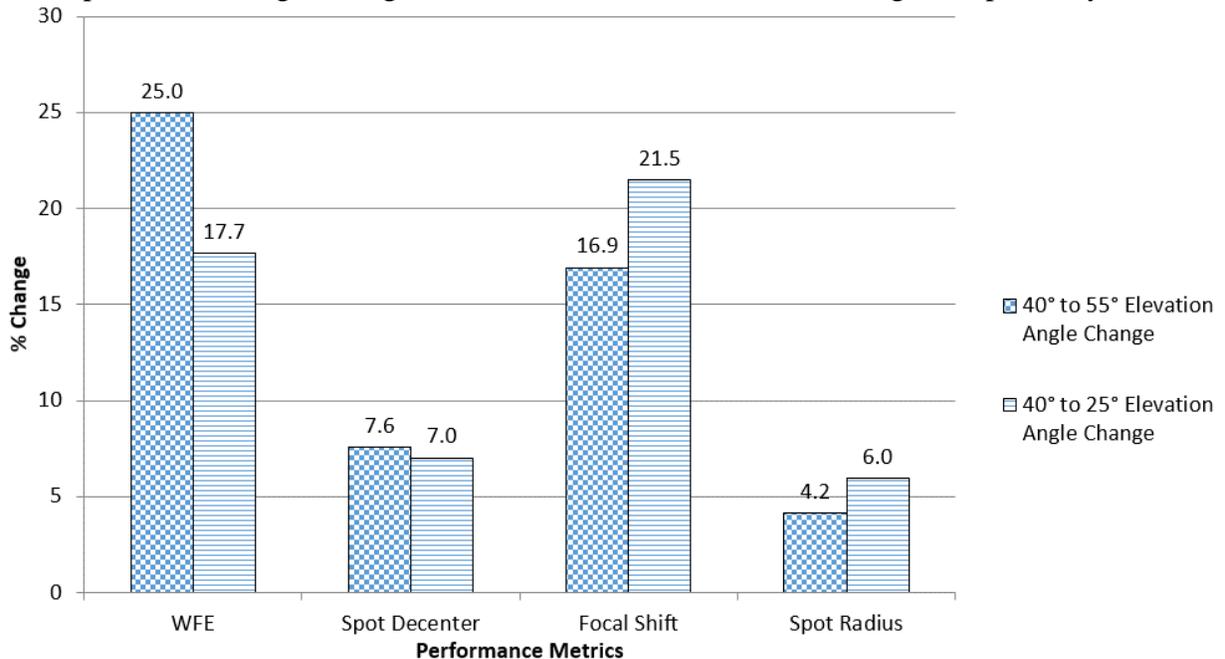

**Figure 37.** Percent change in performance metrics due to telescope elevation angle change. The larger the percentage, the more sensitive that metric is to elevation angle change. Absolute values used.

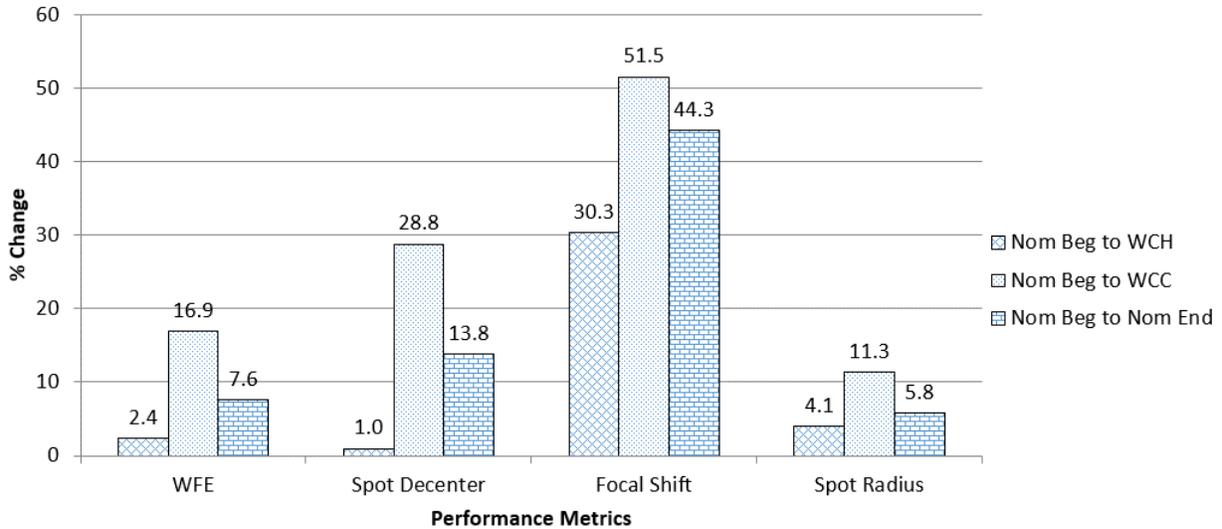

**Figure 38.** Percent change in performance metrics due to thermal scenario change. The larger the percentage, the more sensitive that metric is to thermal scenario change. Absolute values used.

The following is an example for interpreting Figure 37. This figure only shows the impact telescope elevation angle change has on optical performance. This figure assumes that the 40° elevation angle is the nominal position of the telescope. These percentages thus represent the change in elevation angle *from* the telescope's nominal position. Each percentage shown is an average change across the four thermal scenarios. For example, the 25.0% wavefront error change is an average of the four thermal scenarios when the telescope elevation angle changes from 40° to 55°. The Nominal Beginning of night showed a 27% wavefront error increase. The Worst Case Hot showed a 28% wavefront error increase. The Worst Case Cold showed a 24% wavefront error increase. The Nominal End of Night showed a 21% wavefront error increase. Averaging these together yields a 25% increase in wavefront error when the telescope elevation angle changes from 40° to 55°. The intent is to remove the effects the thermal scenarios have on the performance metrics. The same interpretation should be used for the remainder of Figure 37. It should be noted that for the focal shift, 0mm of shift occurs when the telescope is at a 0° elevation angle during ground alignment. The average focal shift at a 40° elevation angle is 2.1mm, so the percentage change is from that value.

Figure 38, on the other hand, only shows the impact of thermal scenarios on optical performance. This figure assumes that the nominal beginning of night is the nominal thermal scenario. These percentages thus represent the change in thermal environment *from* the nominal thermal environment. Each percentage shown is an average change across the three telescope elevation angles. For example, the 30.3% change in focal position is an average of the three elevation angles when the thermal environment changes from nominal beginning of night to worst-case hot.

These figures show that the impact on optical performance was significant for both telescope elevation angle and thermal scenario changes. The wavefront error was more sensitive to the elevation angle changes, with a max change of 25% from nominal, compared with a max change of 16.9% for the changing thermal scenarios. The spot decenter was more sensitive to the changing thermal scenarios, with a max change of 28.8% from nominal, which is much larger than the max change of 7.6% for the change in elevation angles. The change in focal shift was more sensitive to the changing thermal scenarios, with a max change of 51.5%, compared to a max change of 21.5% for elevation angle change. Lastly, the change in spot radius was more sensitive to the thermal scenario change, with a max change of 11.3%, which is larger than the max change of 6.0% for elevation angle.

*4.3 Optical Error Budget*

The overall wavefront error in the system is a combination of the STOP analysis predictions based on thermal and mechanical conditions in addition to the alignment, refocusing, and thermal gradients that were not considered in this analysis. In order to capture these terms without drastically expanding the scope of the STOP analysis, bounding cases were established for each and were added into the final system error budget for wavefront error and decenter using the root sum of their squares. Both budgets are seen for Case 1 in Figure 39.

| | Requirement | CBE | |
|---|---|---|---|
| | 0.143λ RMS | 0.141λ RMS | |
| **Static WFE** | 0.133 | | **Basis for WFE Value** |
| | | 0.108 | Case 1 from STOP |
| WFE generated by STABLE before launch | | 0.074 | Alignment Accuracy Residuals |
| | | 0.022 | Transport Misalignment |
| | | 0.002 | Refocusing Stage Accuracy Residuals |
| | | 0.002 | Thermal Gradients (Structure + M1) |
| **Quasi-Static WFE** | 0.011 | | |
| WFE generated by STABLE over course of night | | 0.011 | WFE due to max temp change over the course of the night |
| **Dynamic WFE** | 0.047 | | |
| WFE generated by STABLE over course of observation | | 0.047 | WFE due to max elevation angle change over course of observation |

**(a)**

| | Requirement | CBE | |
|---|---|---|---|
| | 2620μm | 1669μm | |
| **Static Spot Decenter** | 1603 | | **Basis for WFE Value** |
| | | 899 | Case 1 from STOP |
| Spot decenter generated by STABLE before launch | | 1310 | Alignment of STABLE to star cameras |
| | | 49 | Alignment Accuracy Residuals |
| | | 179 | Thermal contraction of FDM mount |
| | | 105 | Thermal Gradients (Structure + M1) |
| **Quasi-Static Spot Decenter** | 452 | | |
| Spot decenter generated by STABLE over course of night | | 116 | Spot decenter due to max temp change over the course of the night |
| | | 29 | Spot decenter from RFS wander |
| | | 436 | Spot decenter from alignment of RFS when mounted |
| **Dynamic Spot Decenter** | 100 | | |
| Spot decenter generated by STABLE over course of observation | | 79 | Spot decenter due to max elevation angle change over course of observation |
| | | 62 | RFS gravity sag and internal slop due to max elevation angle change over course of observation |

**(b)**

**Figure 39.** STABLE (a) Wavefront Error budget and (b) Spot Decenter budget

As it relates to the wavefront error budget, the optical performance degradation caused by elevation angle and thermal scenario changes make up 83.6% of the total RMS wavefront error at 633nm. The static portion of this wavefront error is the result of the nominal flight elevation angle of 40° and the nominal beginning of night thermal scenario. This accounts for 76.4% of the total RMS wavefront error. The quasi-static portion of the wavefront error budget is the result of the change in thermal environment from nominal beginning of night to nominal end of night. This accounts for 7.6% of the total RMS wavefront error. The Dynamic potion of the wavefront error budget is the result of the telescope elevation angle changing from nominal at 40° to the maximum elevation angle of 55°. This accounts for 32.9% of the total RMS wavefront error.

# 5   Conclusions

High-altitude balloons (HABs) are showing increasing promise as a low cost alternative for doing science-grade data collection. But before they can realize their full potential as science platforms, one of the major technical challenges to address is in characterizing the complex disturbance environment they undergo, including thermal and gravity effects, as well as vibrations generated by system hardware. This disturbance environment was characterized as part of the STABLE Structural, Thermal, Optical Performance (STOP) analysis, which was an integral part of the analysis used to design and verify the system performance. This process contained a number of valuable insights about the subsystem design parameters, modeling techniques, and analysis approaches that improved efficiency for this low-cost ballooning mission.

Determining the thermal conditions is the first step in assessing STABLE's optical performance and was critical to establishing the bounding conditions for the rest of the analysis. An holistic view of the thermal environments and what affects it allow to identify four bounding thermal conditions: nominal beginning- and end-of-the night, worst-case hot, and worst-case cold which were then run through the STOP analysis. This approach simplified the typically complex and slow effort of mapping the thermal model results to the structural analysis model, but represents a much more conservative approach to assessing the thermal performance. The thermal results show that the temperature differences across the beginning and end of a given nominal flight dominate any variation in worst-case hot or cold cases.

The structural portion of the STOP analysis was a critical element in assessing and refining the system design, and many iterations were performed to support various trade studies. The analysis showed that the system can meet its desired optical performance (as suggested by its heritage, testing of the STABLE telescope, and the completed analysis), but the use of contact features to support the primary mirror introduced several complications in the modeling and analysis of the optical system. The uncertainties in the primary mirror mount were minimized by implementing shims with temperature-dependent viscoelastic properties and eliminating contact friction (including altering the flight system by coating the shim surface with Teflon to better match this modeling assumption).  The analysis also showed that one of the biggest contributors of wavefront error was the gravity sag of the primary mirror at the higher telescope elevation angles, which can be improved by light-weighting the mirror. Hardware testing – especially over temperature and flight elevation angles – could further the understanding of the optical-mechanical behavior of the STABLE system and could mitigate risks associated with the uncertainties in the modeling and design.

The optical part of the STOP analysis provided concrete answers about the system's performance against key optical metrics such as wavefront error, decenter, focal shift, and spot size. These demonstrate the sensitivities and challenges associated with maintaining optical performance in the wide range of thermal and gravity conditions experienced on a HAB. STABLE's sensitive optical prescription, which generates 37 µm of system focus shift when the spacing between the primary and secondary mirrors changed by 1 µm, make it particularly difficult to maintain performance across all thermal conditions. Future ballooning missions may consider a less sensitive optical prescription and a thoroughly athermalized telescope structure to bypass some of these concerns. This could be accomplished by using more low CTE components (carbon fiber, Invar, Titanium) or a mechanical design which works to minimize differential contractions between optics based on material properties and expected component temperatures.

An optical error budget was generated which accounted for both the results of the STOP analysis as well as errors associated with alignment, refocusing, and thermal gradients that were not considered in this analysis. The wavefront error and spot decenter error budgets were generated for Case 1, with the contributing sources of error accounted for using the root sum of their squares. The wavefront error of $0.141\lambda$ RMS at 633nm met the requirement of $0.143\lambda$ RMS. The spot decenter of 1669um met the requirement of 2620um.

This STOP analysis brought to light the impact that various design and loading parameters had on optical performance of the system. The first major analysis investigated the effect of thermal gradients on the primary mirror. It was found that based on the geometry of the Zerodur mirror and the thermal environment that the thermal gradients produced only minor changes in wavefront error (<1%). It was thus assumed that bulk soak temperatures could be used for remaining analysis. The preload on the primary mirror was a major design parameter that was analyzed, with 15.6N of preload being the optimal amount. Greater preload generated significant additional system wavefront error. The thermal environment and the elevation angle of the telescope were the largest contributors to the system wavefront error, with 83.6% being generated by the two. It was found that the thermal effects caused significant changes in spacing between the primary and secondary mirrors, which required the refocusing stage to reposition the camera. The thermal scenarios also generated additional wavefront error at each mirror as the mirror assemblies cooled. Thermal control could be used to mitigate these effects but comes with increased system complexity. It was found that the wavefront error was more sensitive to changes in telescope elevation angle from 40° than changes in thermal scenario from the nominal beginning of night case. This was the result of the telescope structure and primary mirror flexing as the gravity vector changes. As telescope elevation angle changes happen on a faster time scale, this was the more concerning of the two contributors. Observing stars lower in the sky would help to minimize the optical performance degradation due to elevation angle, as higher elevations caused more error.

The thermal-structural-optical-performance analysis – performed over twelve gravity and thermal conditions – shows that the STABLE system can achieve the 0.6 Strehl Ratio at its required nominal beginning of night case and across all of the thermal scenarios when at the lowest elevation angle. Because of this analysis, the mission operations can be tuned to better meet the needs of the pointing control subsystem (for example, by targeting stars at higher elevations earlier in the mission). The STABLE system – once flown – is positioned to be a dramatic validation of not only the STOP analysis results but also a meaningful demonstration of the sub-milliarcsecond pointing stability capabilities (and therefore, the high scientific value) of a balloon-based observation platform.

## Appendix A:

*Acknowledgments*

The work described in this paper was carried out at the Jet Propulsion Laboratory, California Institute of Technology, and was sponsored by the NASA JPL Astronomy Directorate through an agreement with the National Aeronautics and Space Administration. The authors would like to thank Jeff Booth at JPL for his continued support for the STABLE project and providing the funding for the write up of this work. The authors would also like to thank John Peacock for his help securing funding for work done at the University of Durham. RJM is supported by the Royal Society.

**Mike Borden** is an optomechanical engineer at the NASA Jet Propulsion Laboratory. He received his BS degree in Mechanical Engineering from the Milwaukee School of Engineering in 2008 and his MS in Optical Science from the University of Arizona in 2012. His current engineering interests include deep space optical communications and astronomical instrumentation. He is a member of SPIE.